\documentclass[twocolumn,referee]{svjour3}          
\smartqed  
\usepackage[utf8]{inputenc}
\usepackage{babel}

\usepackage{stfloats}
\usepackage{graphicx}
\usepackage[hidelinks]{hyperref}
\usepackage{enumitem}
\usepackage{subfig}
\usepackage{float}
\usepackage{newfloat}
\usepackage{booktabs}
\usepackage{multirow}
\usepackage{textgreek}
\usepackage{amsmath}
\usepackage{microtype}
\usepackage{pifont}

\usepackage{wasysym} 

\usepackage{todonotes}
\presetkeys%
    {todonotes}%
    {inline,backgroundcolor=yellow}{}
\usepackage[T1]{fontenc}

\journalname{Preprint}

\begin{document}

\title{Visual Notations in Container Orchestrations:\\ An Empirical Study with Docker Compose}
\titlerunning{Visual Notations in Container Orchestrations: An Empirical Study with Docker Compose}

\author{
  Bruno Piedade
  \and
  João Pedro Dias 
  \and
  Filipe~F.~Correia
}


\institute{
  Bruno Piedade \at
  Faculty of Engineering, University of Porto.\\
  \email{up201505668@fe.up.pt}
\and
  João Pedro Dias \at
  Faculty of Engineering, University of Porto. BUILT CoLAB.\\
  \email{jpmdias@fe.up.pt}
\and
  Filipe F. Correia \at
  Faculty of Engineering, University of Porto. INESC TEC.\\
  \email{filipe.correia@fe.up.pt}
}

\date{Received: date / Accepted: date}

\maketitle
  
\section*{Abstract}

\textbf{Context.} Container orchestration tools supporting infrastructure-as-code allow new forms of collaboration between developers and operatives. Still, their text-based nature permits naive mistakes and is more difficult to read as complexity increases. We can find few examples of low-code approaches for defining the orchestration of containers, and there seems to be a lack of empirical studies showing the benefits and limitations of such approaches.

\noindent \textbf{Goal \& method.}  We hypothesize that a complete visual notation for Docker-based orchestrations could reduce the effort, the error rate, and the development time. Therefore, we developed a tool featuring such a visual notation for Docker Compose configurations, and we empirically evaluated it in a controlled experiment with novice developers.

\noindent \textbf{Results.} The results show a significant reduction in development time and error-proneness when defining Docker Compose files, supporting our hypothesis. The participants also thought the prototype easier to use and useful, and wanted to use it in the future.

\section{Introduction}

The concept of \textit{infrastructure-as-code} (IaC) pertains to the management of infrastructure (\textit{i.e.}, hardware, software and network resources) through configuration files within a code-base~\cite{sousa2018}. Early tools to support this practice focused on bare-metal infrastructure~\cite{pandey2012investigating,Humble2010} but the notion later expanded to the management and provisioning of infrastructure resources on the cloud~\cite{sousa2018,sousa2016}. \textit{Containers}, and Docker in particular, rely on IaC to allow developers to fully specify runtime environments~\cite{merkel2014,Pahl2017}, in a much more lightweight way than virtual machines allow it~\cite{Joy2015,Sousa2015}. Despite the interest given by industry and research to IaC topics in the last few years, there is a consensus that the best practices for developing and maintaining IaC are still weakly established~\cite{kumara2021}. 

\textit{Docker Compose} is a tool for orchestrating multiple containers using Docker. It supports defining an orchestration through a YAML file, by which it configures the containers of the application, the corresponding images and how they are related to each other, the volumes for data persistence, and the networks for connecting the containers. The containers can then be run conventionally through the command-line interface (CLI), resulting in the creation or execution of the declared resources. The YAML files used by Docker Compose follow a well-defined format named \textit{Compose Specification} that claims to be a \textit{``developer-focused standard for defining cloud and platform agnostic container-based applications''}~\cite{composespec}. The Compose Specification resulted from unifying the file formats of versions 2.x and 3.x of Docker Compose. The specification supports the definition of \textit{services}, \textit{networks}, \textit{volumes}, \textit{configs}, and \textit{secrets}.

The process of setting up orchestration files for simple systems is reasonably straightforward, but the textual nature of these files may become challenging as the complexity of the system increases, due to the number or the heterogeneity of the containers. In such cases, we expect that understanding container dependencies becomes difficult, as related definitions begin to get further apart within the file. Also, advanced configuration aspects, such as port mapping and volume management, might be confusing for inexperienced users. Additionally, developing such files by trial-and-error seems to be common \cite{reis2021developing}, and there is some evidence that misconfigurations in IaC scripts are a real concern~\cite{Rahman}. 

\textit{Low-code} approaches to software development enable the visual development of applications, allowing to create software through a graphical user interface rather than the usual text-based computer programming. We find such approaches useful for different purposes and domains, such as in manufacturing, where it is often used for configuring programmable logic controllers (PLC) via ladder and sequential function charts~\cite{bolton2006ladder,torres2020real}. In software engineering, visual notations like the Unified Modeling Language (UML) are reasonably well known and adopted~\cite{ozkaya2020}. More recent applications of visual approaches exist for educational purposes, and in the area of Internet-of-Things (IoT)~\cite{ray2017survey,Dias2018,ihirwe2020low}. There are also examples of such approaches in the operations field, such as for managing cloud and container resources---some of them focusing specifically on Docker technologies~\cite{McKendrick2017}.

We hypothesize that a complete visual approach has the potential to be useful for a broad audience of end-users ranging from first-time developers who benefit from some support in understanding how the technology works to more experienced users who might take advantage of the visualization aspects to have a clearer overview of complex configurations. 

In the work reported in this article, we empirically evaluate a low-code approach for container orchestration. We expand on the work presented at the LowCode 2020 workshop~\cite{piedade2020}, providing a detailed review of the most relevant related works and a more detailed account of the empirical study and of the discussion of the results.

Next, in this article, we start by presenting our research goals and methodology (\textit{cf}. Section~\ref{sec:research-goals}) and discuss relevant related works (\textit{cf}. Section~\ref{sec:related-work}). Then Section~\ref{sec:dockercomposer} describes our approach, and Sections~\ref{sec:empirical-study} and \ref{sec:res} the empirical study and respective results. We end with a discussion of the validation threats and some closing remarks (\textit{cf}.~ Sections~\ref{sec:validationthreats} and~\ref{sec:conclusions}).

\section{Research Goals and Methodology}
\label{sec:research-goals}

The purpose of this work is to explore the benefits of low-code for the development of docker compose orchestration files. In particular, the hypothesis that guides this work is that \textit{a complete visual notation for developing orchestration files can improve the overall developer experience and reduce the error proneness and development time}.

By a \textbf{complete visual notation}, we mean a way to visually express all the elements of an orchestration file that are supported by its text notation. This includes elements such as \textit{containers}, \textit{volumes}, \textit{networks}, \textit{configs}, and \textit{secrets}, as well as different relationships and dependencies.

By \textbf{orchestration files} we mean the descriptions of what containers make a given system, and how they depend on each other and on infrastructure resources. For the scope of this work we consider the Docker Compose Specification~\cite{composespec}.

By \textbf{developer experience}, we mean the overall ease-of-use and intuitiveness of the full experience, considering the steps and actions needed to successfully specify a container orchestration setup.

By \textbf{error proneness} and \textbf{development time}, we respectively mean the number of errors and execution attempts, and the time required to successfully specify a container orchestration setup.

Given this, we consider the research questions:

\begin{enumerate}[label=\textbf{RQ\arabic*},leftmargin=2\parindent]
    \item \textit{To what extent does a visual notation for the orchestration of (Docker) containers reduce the development time?}\\ 
    We aim to understand if a visual notation is truly useful in reducing the time of development of a Docker Compose file. \label{hypothesis:RQ1}
    \item \textit{To what extent does a visual notation for (Docker) orchestration files reduce the number of errors?}\\
    We aim to understand if a visual notation is truly useful in reducing error proneness while orchestrating a Docker Compose file. \label{hypothesis:RQ2}
    \item \textit{What is the perception of developers towards using a visual notation for the orchestration of (Docker) orchestration files?}\\
    We aim to understand if a visual notation is perceived by the developers as enjoyable and useful, and if they show intention of using it again in the future after being exposed to it. \label{hypothesis:RQ3}
\end{enumerate}

To find answers to these research questions we start by \textbf{surveying existing visual approaches for managing and orchestrating container and infrastructure resources} (\textit{cf.}~Section~\ref{sec:related-work}), seeking to find works that come closest to providing insights to our research questions. 

Next, we develop a \textbf{tool prototype} that offers a low-code environment, with a complete visual notation for developing orchestration files for Docker Compose (\textit{cf.}~Section~\ref{sec:dockercomposer}). We name this tool \textit{Docker Composer}, and we use it to \textbf{empirically evaluate} the benefits of using such a visual notation. More specifically, we conduct a controlled experiment with novice software developers, where we gather performance and perception-based metrics, and compare the use of a visual notation with the conventional text-based one (\textit{cf.}~Section~\ref{sec:empirical-study}).

\section{Related Work}
\label{sec:related-work}

As we sought to propose and evaluate a visual approach for developing container orchestration files, we evaluated related works employing visual tools for managing and orchestrating containers. These tools allow to define or inspect Docker resources, locally or remotely, usually for development purposes. Such resources can include \textit{containers}, \textit{images}, \textit{volumes}, and \textit{networks}. Many times, they work as wrappers for Docker's CLI commands, from common functions such as container creation and deletion to the orchestration of containers.

Albeit less thoroughly, we also review tools to visually manage and orchestrate infrastructure, as they address a closely related domain and could support useful additional insights. Such infrastructure resources can include physical (\textit{i.e.}, hardware and facilities), virtual machines, and network resources. 

We started our analysis by using \textit{Google Scholar} and, upon realizing the low number of research works addressing visual approaches for container and infrastructure resources, extended our review to other visual tools that are readily available to practitioners by querying \textit{Google} and \textit{GitHub}. We based the analysis of these tools primarily on their available documentation and manually installed them when necessary. 

\subsection{Visual Tools for Managing and Orchestrating Containers and Infrastructure}

The next paragraphs briefly describe the tools that we have surveyed. We sought to identify capabilities and limitations of these tools and, for those supporting the management and orchestration of containers, we assessed the extent to which they support a \textit{complete visual notation} (\textit{cf}. Section~\ref{sec:research-goals}).

\textbf{DockStation\footnote{DockStation, available at \url{https://dockstation.io/}}} seems to be considerably adopted, with over 1.8K \textit{stars} on GitHub as of January of 2022. It provides a native GUI for handling Docker containers in local and remote environments and is aimed primarily at development. It supports container management, such as creating and deleting containers, and a few container monitoring utilities, including performance graphs. 

\begin{figure*}[t]
  \begin{center}
    \leavevmode
    \includegraphics[width=.8\textwidth]{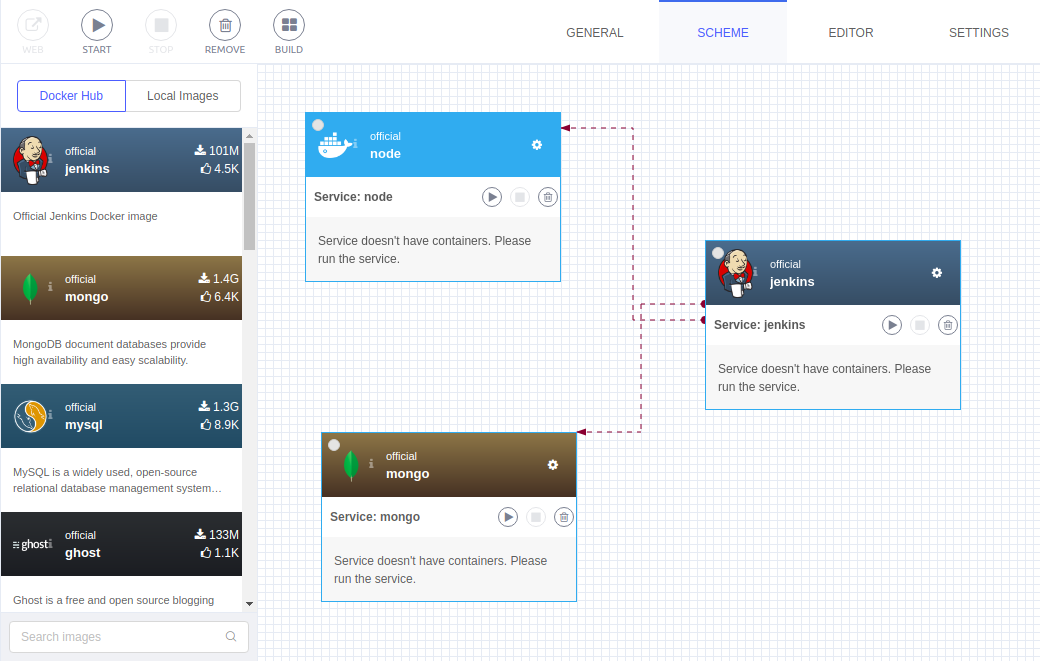}
    \caption{A simple \textit{docker-compose} configuration in DockStation.}
    \label{fig:dockstation}
  \end{center}
\end{figure*}

We can instantiate a new \textit{project} by creating a new or loading an existing \texttt{docker-compose.yml} file. We then visualize the overall containers scheme. The containers and relationships are represented in a graph-like diagram, and we can edit different aspects through form fields, such as environment variables, volumes, and ports. Changes are reflected in an underlying \texttt{docker-compose.yml} file, which can also be seen in its text form within the editor, in a different tab. In the same way, changes to this text form will be reflected on the diagram.

Fig.~\ref{fig:dockstation} showcases the \textit{scheme} perspective in the \textit{project} tab for a simple Compose orchestration file and some additional UI elements. 
The top action bar includes quick actions which trigger Docker Compose commands such as \texttt{docker-compose up} for the start button. In the scheme itself, the boxes represent the containers and the dotted arrows represent the \texttt{depends\_on} relation between the containers. To add a container to the scheme, the user can drag the intended image from the palette, to the left, to the scheme area, to the right. In the current version, it is not possible to visually add dependencies between containers (\textit{i.e.}, \texttt{depends\_on}), requiring the user to instead use the editor and add the dependency textually in the \texttt{docker-compose.yml} file.

This tool stands as one of the closest to the one we propose in this work, although a few limitations stand out: (a)~volumes are specified via a form and have no visual notation; (b)~dependencies are represented visually but they can only be defined through the text editor; (c)~configs and secrets are not represented visually and can only be defined through the text editor; and (d)~both visual and textual representations are available within the tool but one cannot view or interact with both simultaneously. The user has to explicitly click on a save button in the editor for the changes to persist before switching.

\textbf{Admiral\footnote{Admiral, by VMware, available at \url{https://github.com/vmware/admiral}}} seems to have a smaller user base than DockStation, with over 252 stars on GitHub as of January 2022\footnote{As of January of 2022 the project is marked as \textit{archived} on GitHub, suggesting that no future developments are to be expected.}. It offers a web-based GUI for container management and provisioning over a cluster of infrastructure. Unlike DockStation, this tool is mainly deployment and production-oriented. Besides the provisioning of single containers, it supports orchestrating containers by the definition of \textit{templates}. These include four main components: \textit{containers}, \textit{volumes}, \textit{networks}, and \textit{closures}\footnote{\textit{Closures} are a notion supported by VMWare tools, and not an official feature of the Docker Compose specification.}. Each can be created and configured via a form-based user interface. It is then possible to visually connect each container or closure with a network or volume by click-and-dragging the mouse pointer from the source to the target component. Each template can be directly provisioned to a configured cluster or exported in one of two formats---YAML Blueprints\footnote{\textit{YAML Blueprints} is a format specified by VMWare and used by tools provided by the company, such as \textit{vRealize Automation}.} and Docker Compose files---and it is also possible to import from these file formats to visualize and edit the orchestration.

\begin{figure*}[t]
  \begin{center}
    \leavevmode
    \includegraphics[width=.8\textwidth]{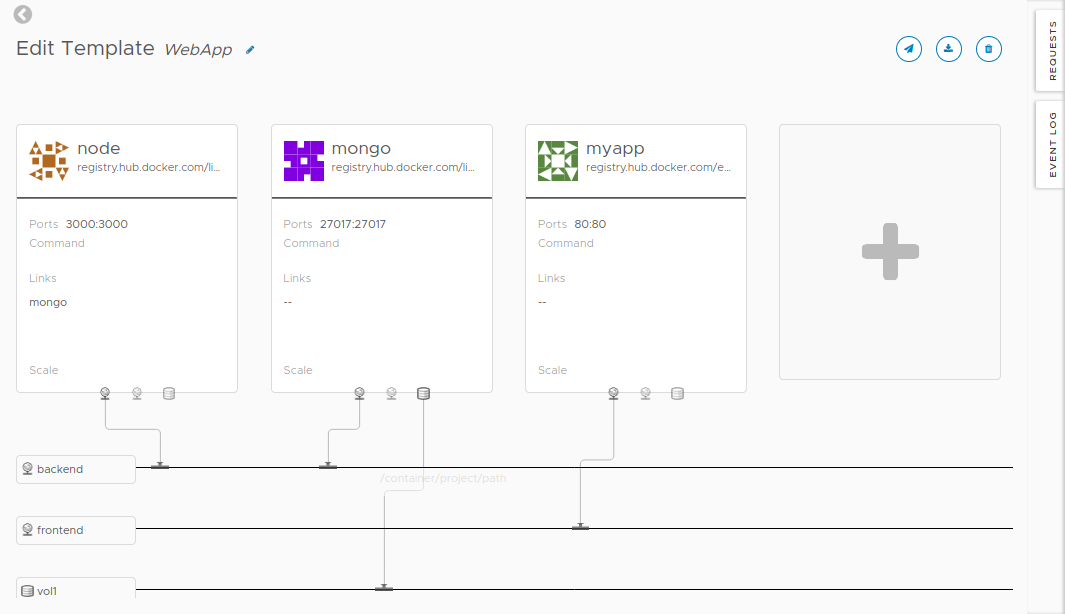}
    \caption{Sample of Admiral's template visual orchestrator.}
    \label{fig:admiral}
  \end{center}
\end{figure*}

Fig.~\ref{fig:admiral} displays a simple template containing 3 containers, 2 networks, and 1 volume. The user can add a new component (container, network, volume, or closure) by hovering over the empty box with the plus icon and clicking on the desired element. Upon which they are redirected to the corresponding form to edit its properties. Each container has a set of network and volume anchor points, located at the bottom, for the total number of networks and volumes declared in the configuration (3 in this instance). These allow the user to attach the containers to their corresponding volumes and networks. Dependencies between containers, known as links, are displayed and directly editable as properties of a container, within its box. To edit more advanced properties, the user must expand the container and access its full edit form.

Four limitations stand out in Admiral: (a)~like in DockStation, configs and secrets cannot be visually represented; (b)~the dependencies between containers (\texttt{depends\_on} and \texttt{links}) can be specified via a drop-down, but are not represented as lines connecting the containers; (c)~the user is not allowed to rearrange any of visual elements, with the exception of network connections; and (d)~some elements supported by Admiral are specific to VMWare tools (\textit{e.g.}, closures) but not part of the Docker Compose specification, and they are lost (with no warning) when exporting an orchestration to a \textit{docker-compose.yml} file, possibly giving a false sense of what developers may expect to be actually building with the tool.

\begin{figure}[h]
  \begin{center}
    \includegraphics[width=1\linewidth]{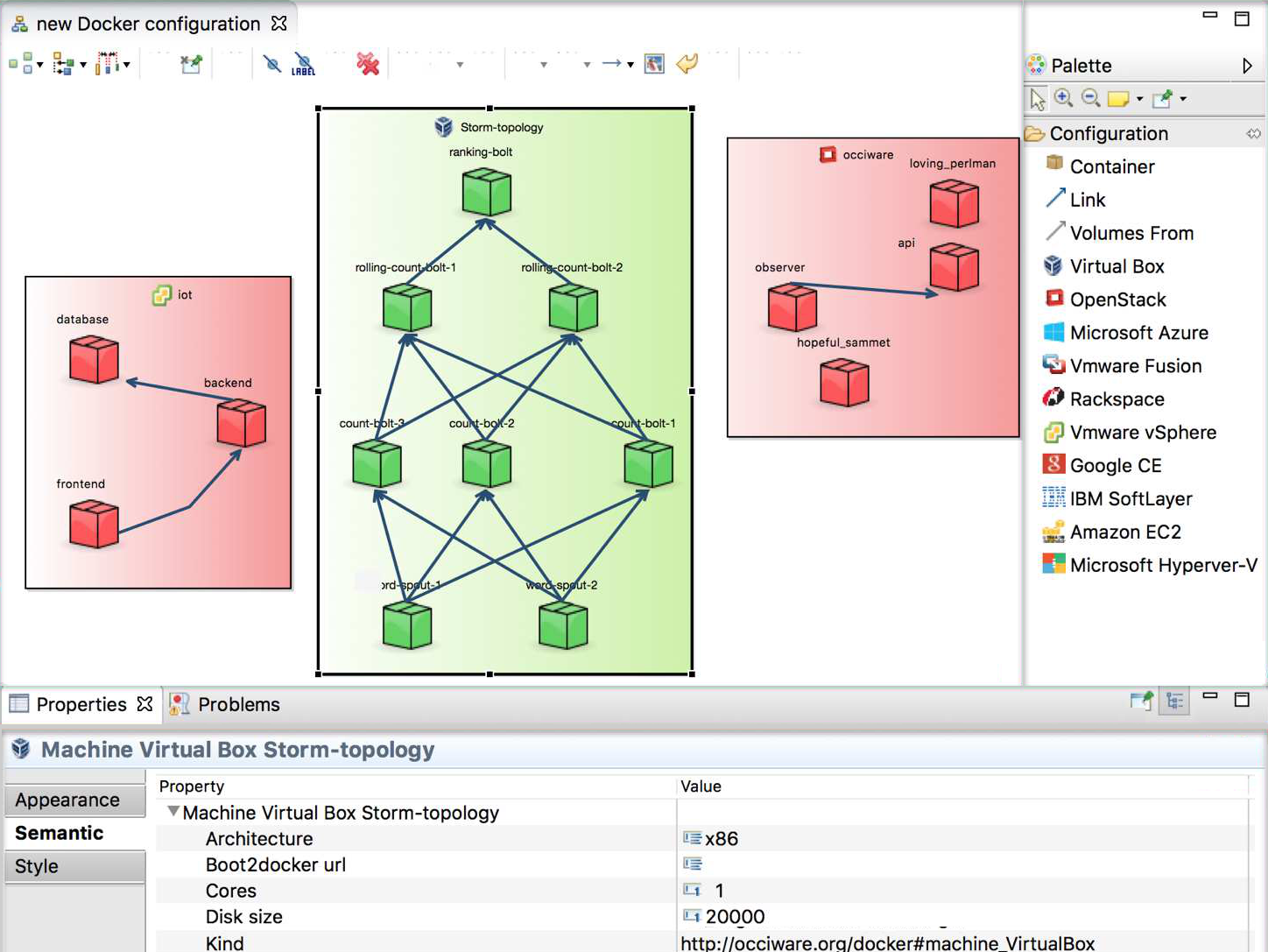}
    \caption{Sample of Docker Designer's user interface, adapted from Paraiso \textit{et al.}~\cite{Paraiso2017}.}
    \label{fig:dockerdesigner}
  \end{center}
\end{figure}

\textbf{Docker Studio}\footnote{Docker Studio is available at \url{http://occiware.github.io/content/user-guides/snapshot/connector-docker.html}.} and its predecessor, \textit{Docker Designer}~\cite{Paraiso2017}, do not seem to be in common use by professionals. The tool is an Eclipse-based prototype, offering a native GUI. It employs a model-driven approach to address deployment and maintenance in production environments. Its user interface is shown in Fig.~\ref{fig:dockerdesigner}. It features a palette on its right side that allows configuring different container and infrastructure elements. To the center, there is a design area that provides a graphical representation of model. The tool allows to visually establish the dependencies between containers and represent their volumes and networks. Docker Studio allows also to run or stop containers in their execution environment. Green and red colors represent respectively resources that are in \textit{running} or in \textit{stopped} states.

Limitations of this tool include: (a)~like in DockStation and Admiral, \textit{configs} and \textit{secrets} cannot be defined visually; and (b)~there is no way of accessing Docker's output, which can become a concern when trying to troubleshoot any issue that might arise.

\begin{figure}[h]
  \begin{center}
    \leavevmode
    \includegraphics[width=1\linewidth]{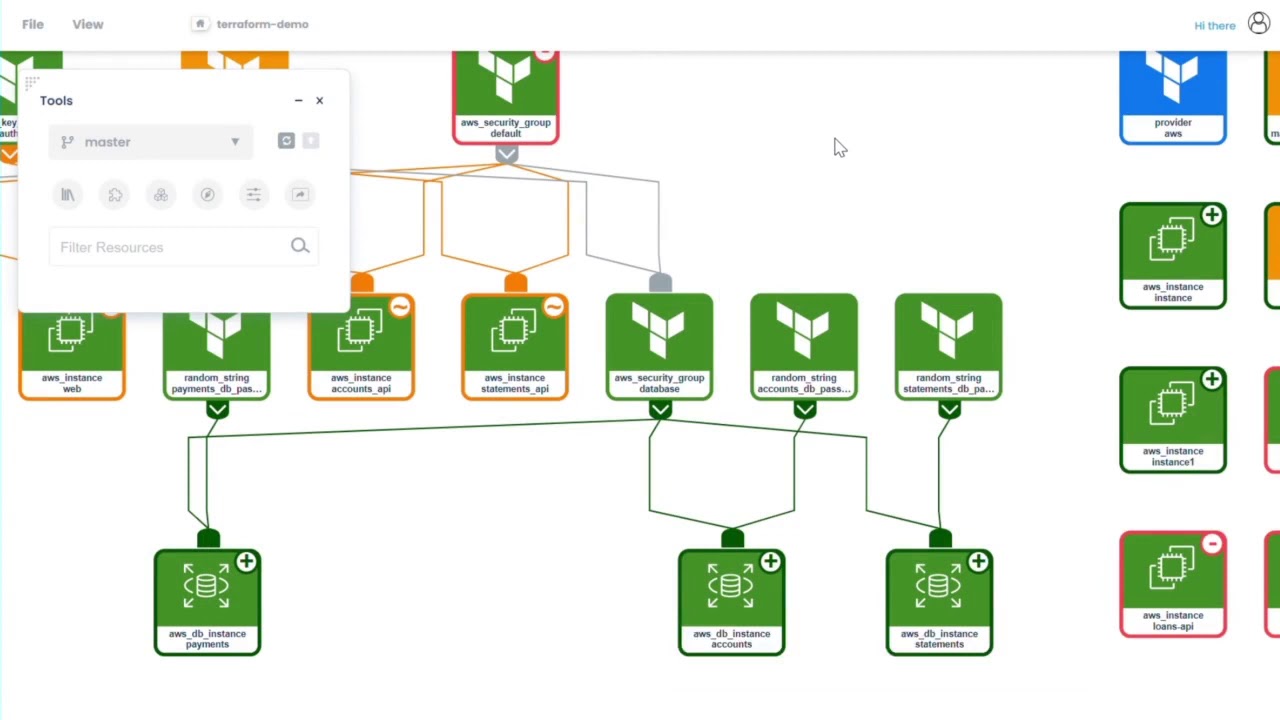}
    \caption{Sample of CodeHerent's user interface.}
    \label{fig:code_herent}
  \end{center}
\end{figure}

\textbf{CodeHerent\footnote{CodeHerent, available at \url{https://codeherent.tech/home}}} is a web-based visual development environment, leveraging a hybrid visual programming language (VPL) for editing and visualizing Terraform configuration files. Unlike DockStation, Admiral, and Docker Studio, it does not address container resources. A sample of its user interface can be seen in Fig.~\ref{fig:code_herent}. Although this tool initially adopted a box-based representation in which the different elements are hierarchically organized in boxes, more recent releases opt for a graph-based diagram to represent the distinct elements and their relationships. 

~\\

\begin{figure}[h]
  \begin{center}
    \leavevmode
    \includegraphics[width=1\linewidth]{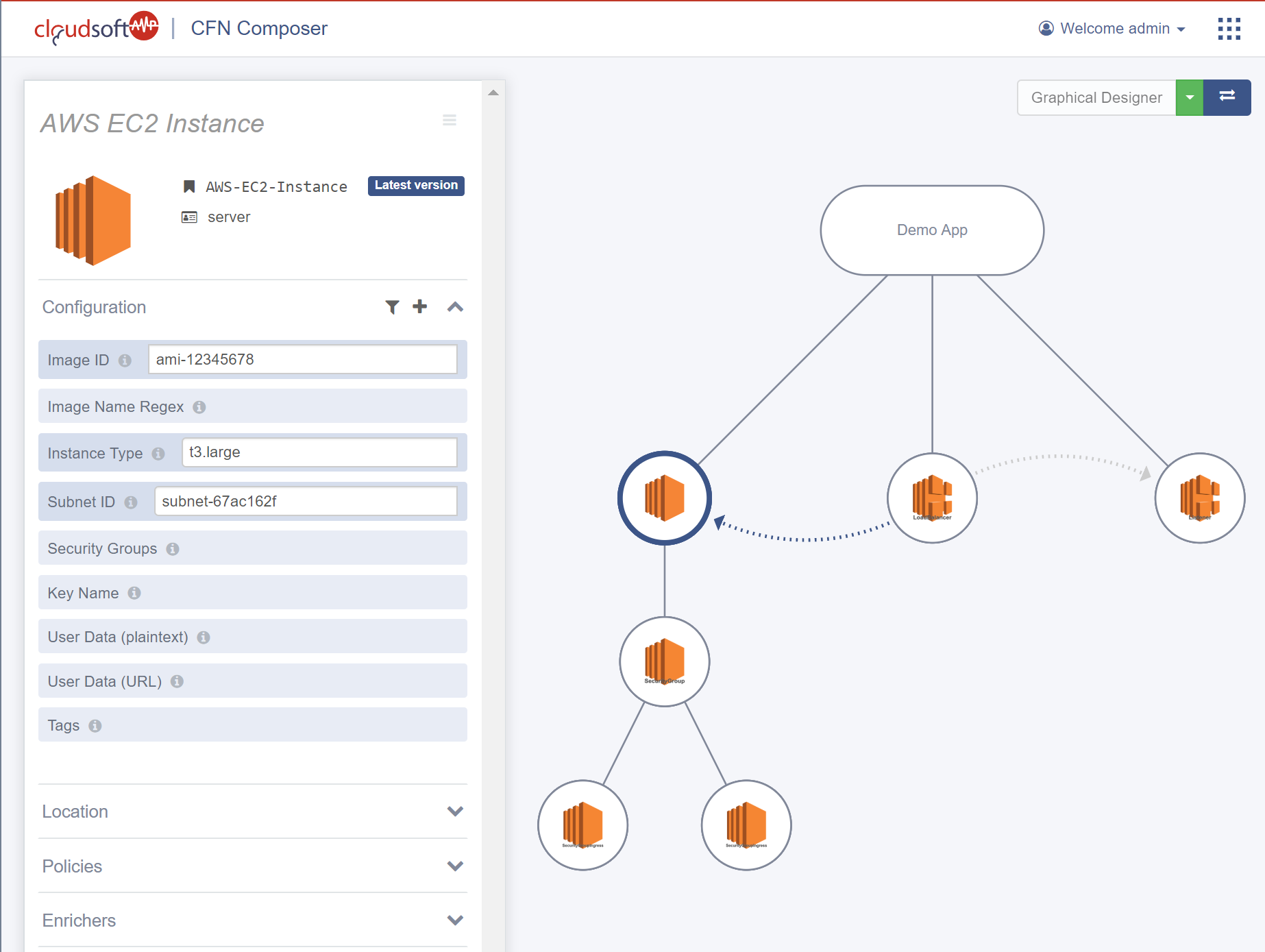}
    \caption{Sample of Visual Composer's interface.}
    \label{fig:cfn_composer}
  \end{center}
\end{figure}

\textbf{Visual Composer\footnote{Visual Composer, by CloudSoft, originally available at \url{https://cloudsoft.io/software/cfn-composer/} and on the AWS marketplace at \url{https://aws.amazon.com/marketplace/pp/prodview-pqc3effdvhy3s}, seems to have been discontinued as of January 2022.}} resembles CodeHerent but focuses specifically on AWS EC2 CloudFormation templates. It uses a web-based GUI, representing infrastructure artifacts following a tree-like diagram and offering multiple types of connections, such as arrows for dependencies and references between resources. The user can add a node as a descendent of another and edit its properties in a form-based interface. Furthermore, it includes snippets of documentation directly accessible by hovering help icons for each element. Similar to other hybrid visual approaches, it supports switching between the visual composer and a built-in textual editor of the corresponding YAML file. A sample of Visual Composer's GUI can be seen in Fig. \ref{fig:cfn_composer}.

\newpage

\begin{figure}[b]
  \begin{center}
    \leavevmode
    \includegraphics[width=1\linewidth]{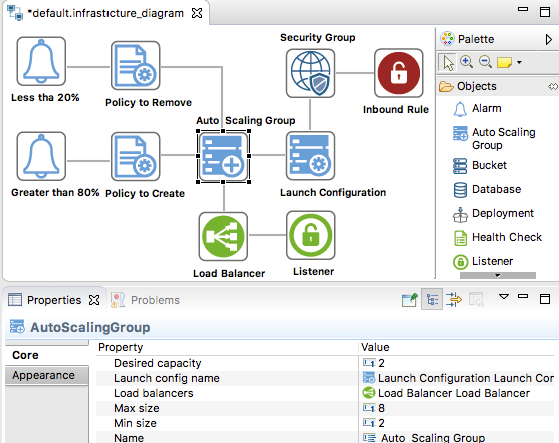}
    \caption{Sample of Argon's interface, adapted from Sandobalin \textit{et al.}~\cite{Sandobalin2017}.}
    \label{fig:argon}
  \end{center}
\end{figure}

\textbf{Argon} is an infrastructure modeling tool supporting different IaC platforms~\cite{Sandobalin2017}. Like Docker Studio, it is an Eclipse-based tool and offers a native GUI. Argon allows selecting such resources from a palette window and editing each resource's properties through a form-based interface. The resulting visually-created model of the infrastructure can then be used to generate scripts for different IaC platforms (\textit{e.g.}, Ansible and Terraform), unlike \textit{Codeherent} and \textit{Visual Composer}, which are bound to specific technologies. The experiments conducted by Sandobalin \textit{et al.}~\cite{Sandobalin2020} with 67 Computer Science students have empirically compared Argon with a well-known IaC tool (Ansible) and showed promise in the effectiveness and perceived ease of use and usefulness of Argon's visual approach.

\subsection{Discussion} \label{sota:services_summary}

We can classify the surveyed tools as simultaneously \textit{form-based} (\textit{i.e.}, using form fields or a spreadsheet-like user interface) and \textit{hybrid} (\textit{i.e.}, combining text and visual systems), according to the scheme proposed by Boshernitsan \textit{et al.}~\cite{Boshernitsan2004}. The tools that come closer to supporting our goals are \textit{DockStation}, \textit{Admiral} and \textit{Docker Studio}, but they have yet to fully explore the potential of a visual approach for the following reasons:

\begin{itemize}
    \item \textbf{Incomplete visual notations.} One of the issues found is the lack of visual representations for some of the elements supported in the Docker Compose Specification. In particular, Admiral and Docker Studio appear to be the most complete, but they do not support specifying Docker \textit{configs} and \textit{secrets}. 
    An incomplete visual notation encourages developers to fall back into the text notation when something cannot be understood from the visual notation. We believe that the friction caused by this additional context-switching may discourage using the visual notation or reduce the benefits to be gained from using it.
    
    \item \textbf{Limited visual editing.} Only Docker Studio allows editing all of the elements and properties of Docker Compose that it supports (\textit{configs} and \textit{secrets}, as stated before, are not supported at all). DockStation allows to circumvent this limitation by providing a textual editor for the \textit{docker-compose.yml} file within the application, but the resulting workflow does not provide a streamlined experience; a user must switch between the textual and visual perspectives as needed.
    
    \item \textbf{Sub-optimal directness.} Current solutions present lower directness~\cite{Burnett1999} than desired, as they require several steps to manipulate a Compose orchestration file visually. For instance, Admiral requires a few navigation steps to create or edit an artifact. The user must first click on a button that leads to a new page with form fields for input. After altering the definition, the user must confirm their action for it to take effect. This indirectness is inconvenient, hindering the workflow, and hides useful information that could otherwise be always visible.
\end{itemize}

Table~\ref{tab:tools} (p.\,\pageref{tab:tools}) summarises the surveyed tools, illustrating their capabilities and limitations in visually representing a container orchestration. This table also contrasts the surveyed tools with \textit{Docker Composer}, a tool that we describe in the next section (Section~\ref{sec:dockercomposer}).

Finally, it is worth noting that, as far as we know, the only one of these tools that was used in the context of an empirical study is Argon~\cite{Sandobalin2020}. The positive results of this study, albeit in the domain of \textit{infrastructure}, encouraged us further to evaluate the benefits and possible limitations of using a low-code approach for orchestrating \textit{containers}. 

\section{The \textit{Docker Composer} tool}
\label{sec:dockercomposer}

We have considered using tools such as \textit{DockStation}, \textit{Admiral} or \textit{Docker Studio} to empirically evaluate the benefits and limitations of a visual notation in the development of container orchestration files. However, we thought it essential that the tool handled all the elements that one can understand or express through the text notation of Docker Compose files. Unfortunately, these tools do not yet provide such support (\textit{cf.} Section~\ref{sota:services_summary} and Table~\ref{tab:tools}), so we have ultimately decided to develop a new tool, which we named \textit{Docker Composer}.

\newcommand{\rotbox}[1]{\rotatebox{45}{#1}}

\begin{table*}[ht]
\centering
    \caption{Comparative overview of visual tools for managing and orchestrating container and infrastructure resources. The latter refer to virtual machines and other resources that can be used to host containers but are not themselves containers or part of containers. A filled circle ($\CIRCLE$) is used for elements that can be diagrammatically represented and edited (\textit{e.g.}, using boxes and arrows); a dot ($\bullet$) for when editing the element is done through a form field; and an empty circle ($\Circle$) for when editing an element requires interacting directly with the fully textual form of the Compose orchestration file.}
    \label{tab:tools}
\begin{tabular}{llcccccclllc}
                &  & \multicolumn{8}{c}{Container Specification}                                                                                                                                                                                                                                                                                     &  & Infrastructure \\ 
                &  & \rotbox{Services} & \rotbox{Volumes} & \rotbox{Networks} & \rotbox{Configs} & \rotbox{Secrets} &  & \rotbox{\texttt{depends\_on}} & \rotbox{\texttt{links}} &  &                \\ \cline{3-10} \cline{12-12}
DockStation     &  & $\CIRCLE$                          & $\bullet$                         &                                    &                                   &                                   &  & $\Circle$                                                      & $\Circle$                                                &  &                \\
Admiral         &  & $\CIRCLE$                          & $\bullet$          & $\bullet$                          &                                   &                                   &  &                                                                & $\bullet$                                                &  &                \\
Docker Studio   &  & $\CIRCLE$                          & $\CIRCLE$                         & $\CIRCLE$                          &                                   &                                   &  & $\CIRCLE$                                                      & $\CIRCLE$                                                &  & $\CIRCLE$      \\
CodeHerent      &  &                                    &                                   &                                    &                                   &                                   &  &                                                                &                                                          &  & $\CIRCLE$      \\
Visual Composer &  &                                    &                                   &                                    &                                   &                                   &  &                                                                &                                                          &  & $\CIRCLE$      \\
Argon           &  &                                    &                                   &                                    &                                   &                                   &  &                                                                &                                                          &  & $\CIRCLE$      \\
Docker Composer &  &    $\CIRCLE$                                &       $\CIRCLE$                            &          $\CIRCLE$                          &             $\CIRCLE$                      &               $\CIRCLE$                    &  &  $\CIRCLE$                                                              &         $\CIRCLE$                                                 &  &       \\ \cline{3-10} \cline{12-12}
\end{tabular}

\end{table*}

\begin{figure}[h]
  \begin{center}
    \includegraphics[width=1\linewidth]{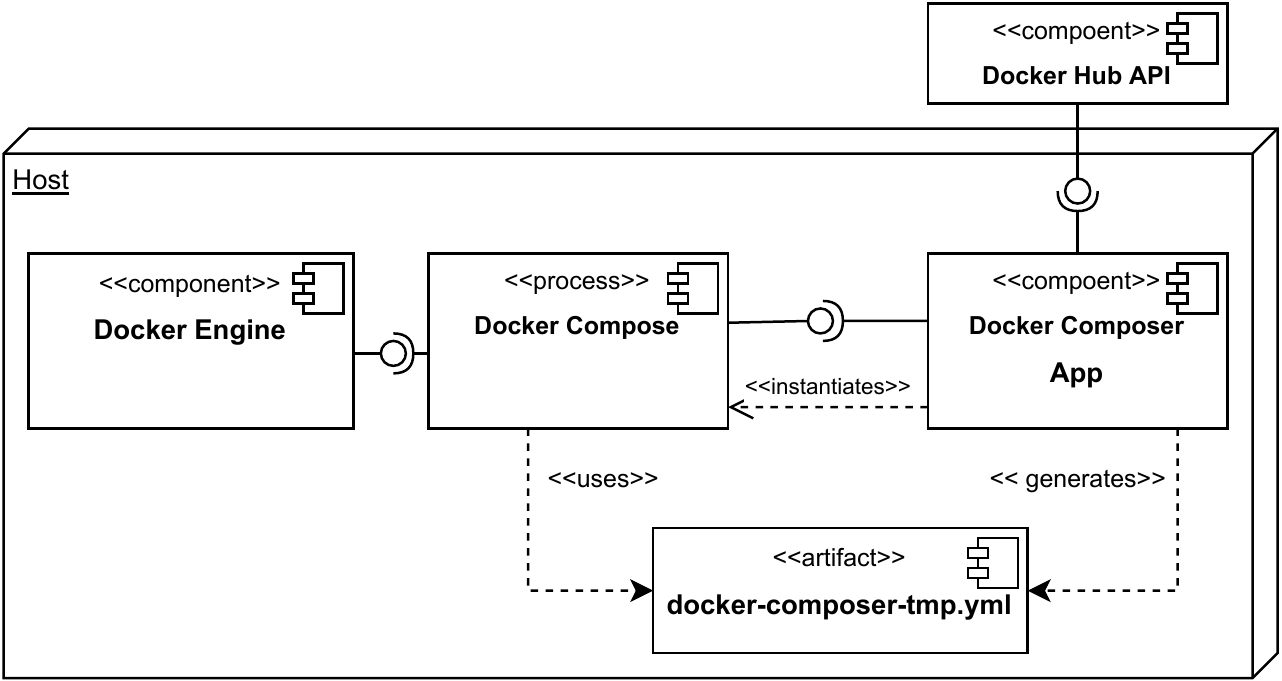}
    \caption{Deployment diagram of the prototype.}
    \label{fig:deployment_diagram}
  \end{center}
\end{figure}

Fig.~\ref{fig:deployment_diagram} shows the high-level architecture of the tool. Within the host environment, the prototype (in the figure, represented as the \textit{Docker Composer App}) generates Docker Compose YAML files and launches shell instances where it executes Docker Compose via CLI commands. In turn, Docker Compose communicates with the Docker Engine. Docker Composer also generates requests to the remote Docker Hub's public API, to receive information about the images hosted on this service. 

Furthermore, Docker Composer can open and save any \texttt{docker-compose.yml} file. Opening one such file translates its contents to an object model, which then supports the features related to visualization and user interaction. This object model is designed to express all the elements that we can find in a Docker Compose orchestration file. Saving back to a file is the reverse process of serializing this object model to a YAML file following the Docker Compose Specification.

In Fig.~\ref{fig:prototype_layout}, we present the prototype's main view, which features five distinct panels, namely:

\begin{figure*}[ht]
  \begin{center}
     \includegraphics[width=0.8\linewidth]{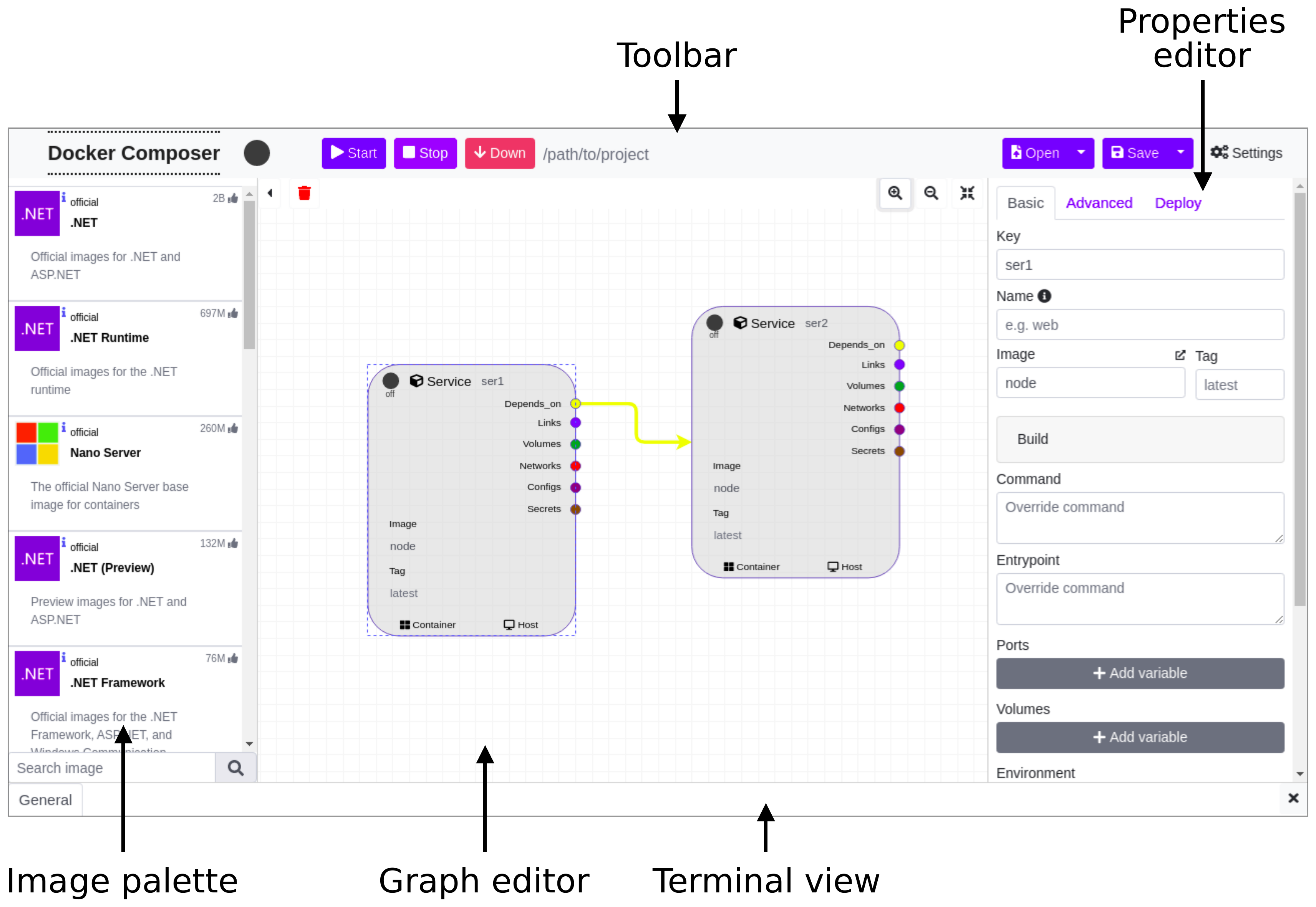}
    \caption{Layout of the prototype's main view, showcasing the graph editor (drawing canvas), main control toolbar, image palette (from Docker Hub) and the properties editor (corresponding to one of the services in the canvas).}
    \label{fig:prototype_layout}
  \end{center}
\end{figure*}

\begin{itemize}
     \item \textbf{Toolbar.} To the left, it includes a status indicator, which lights up different colors according to the state of the running orchestration, and a set of buttons to start and stop the services (containers). To the right, it includes a few buttons for file management. The settings menu allows to set the working directory and adjust preferences when exporting files.
    \item \textbf{Image Palette.} Allows searching for images hosted on Docker Hub and the addition of new services by clicking and dragging the target image and dropping it in the graph editor area. 
    \item \textbf{Graph editor.} This area displays an interactive visual map of the orchestration containing the various artifacts that comprise it and their dependencies.
    \item \textbf{Properties Editor.} Useful to access and edit the various properties of the currently selected object (artifact or connection) in the graph editor.
    \item \textbf{Terminal view.} Displays the output produced by the services (containers) once created and started. It contains a \textit{General} tab with the combined output of all services and additional logs (\textit{i.e.}, Docker Compose logs) and individual tabs for the output of services that comprise the orchestration.
\end{itemize}

\begin{figure}[h]
  \begin{center}
    \leavevmode
    \includegraphics[width=0.78\linewidth]{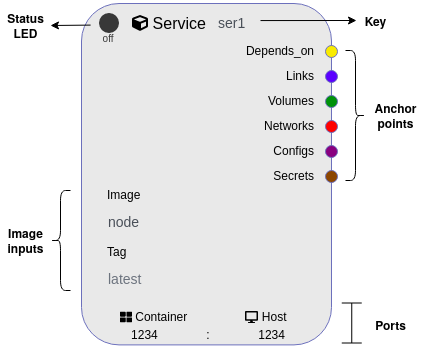}
    \caption{Visual representation of a service artifact node.}
    \label{fig:prototype_artifacts_service}
  \end{center}
\end{figure}

Fig.~\ref{fig:prototype_artifacts_service} shows the visual notation of a service node. It includes a set of anchor points located on the right edge. Each anchor point is used as the source point to set connections between the service and some target artifact. This can be achieved by left-click dragging from the source point to a compatible target artifact. These connections are typed, meaning that only certain artifacts are expected as targets, and the tool only allows this type of connection. To make the type of the connection more explicit, the colors of the anchor points match that of the allowed type of artifact, except for \textit{depends\_on} (yellow) and \textit{links} (blue) anchors. These last two anchors are used to connect services; \textit{depends\_on} establishes the order of container creation, while \textit{links} allows containers to be reachable at an alias hostname.

\begin{figure}[!h]
  \begin{center}
    \leavevmode
    \includegraphics[width=0.95\linewidth]{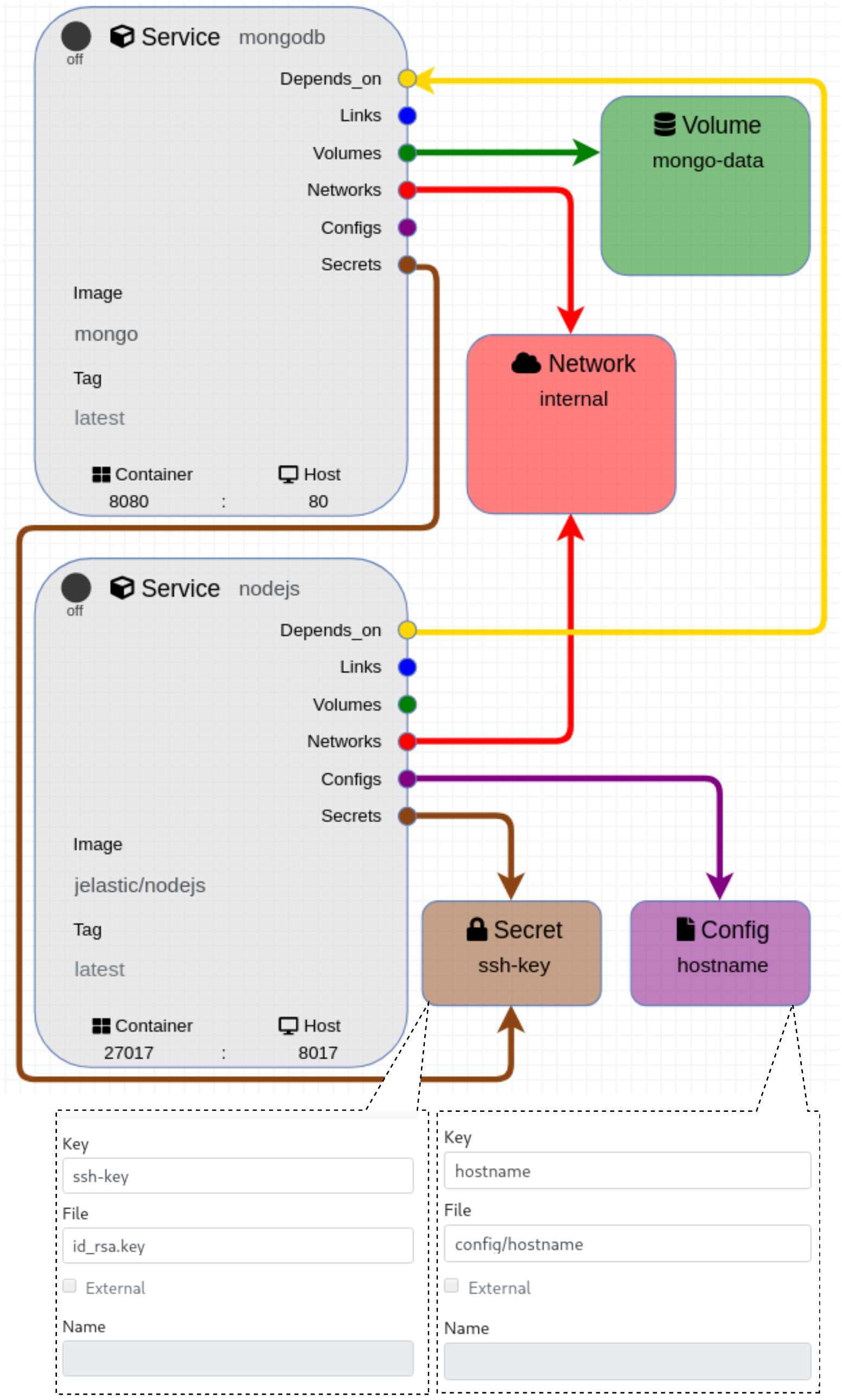}
    \caption{Example orchestration in Docker Composer. The two services share the \texttt{internal} network, and the \texttt{mongodb} service uses a volume for storing its data named \texttt{mongo-data}. Both services also share a secret (an SSH key) and the \texttt{nodejs} container uses a specific \texttt{hostname} config. The two forms to the bottom of the figure show the input fields for editing the \textit{secret} and \textit{config} as they will appear on the right sidebar of Docker Composer when the respective element is selected.}
    \label{fig:prototype_artifacts}
  \end{center}
\end{figure}

Fig.~\ref{fig:prototype_artifacts} shows the remaining elements (\textit{volumes}, \textit{networks}, \textit{configs} and \textit{secrets}). They are represented by a similar notation, only differing in color, size, and labels, depending on their type. All nodes allow to input their key as exemplified in the figure for \textit{secrets} and \textit{configs}.

\begin{figure}[h]
  \begin{center}
    \leavevmode
    \includegraphics[width=1\linewidth]{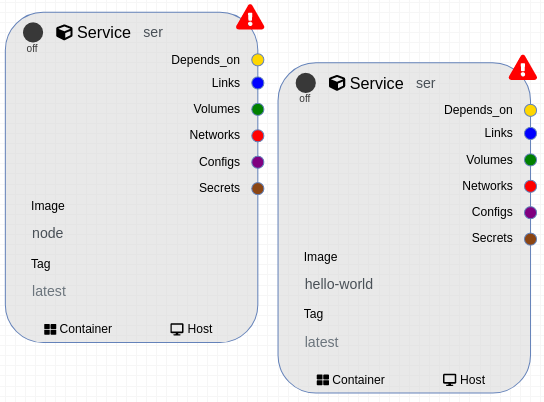}
    \caption{Example of the static validation notation. Both services include the warning icon because they define the same key (\textit{ser}).}
    \label{fig:validation_example}
  \end{center}
\end{figure}

The tool provides static validations while editing an orchestration. These include duplicate key detection and invalid property value formats (\textit{e.g.}, for values specified as time duration or memory size). The result of the validations is conveyed to the users through warning icons that appear near the artifacts' visual representation. It is possible to hover these icons with the mouse pointer to visualize a full summary of the warnings. These inconsistencies are purely presented as warnings and are not enforced as errors and ultimately provide additional feedback to users.
Fig.~\ref{fig:validation_example} shows an example of static validation. The warning in this example results from the use of the same key (\textit{ser}) for both services.

Another form of validation is the mechanism used to control the consistency of some property values. In particular, when defining port mappings, the user cannot define host ports without first setting a container port. We achieve this by controlling whether inputs are disabled or not. Additionally, the connections between containers are \textit{typed}, thus erroneous connections are not allowed, such as trying to connect a \textit{Networks} gate to a \textit{Service} block).

\begin{figure*}[hb]
  \centering
  \subfloat[Docker Compose.\label{fig:docker-compose-spec}]{%
    \includegraphics[width=.24\linewidth]{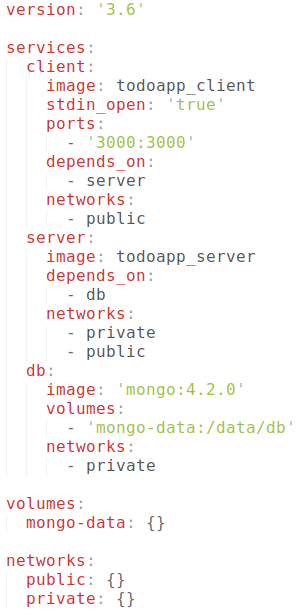}%
  }\hspace{0.8cm}
  \subfloat[Docker Composer.\label{fig:docker-composer-spec}]{%
    \includegraphics[width=.68\linewidth]{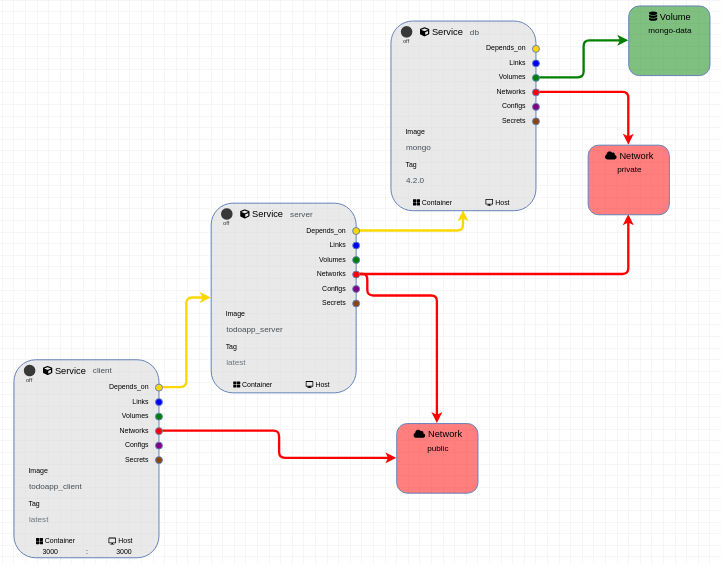}%
  }
  \caption{Concrete example of a Docker compose file. (a)~presents the default textual representation of a \texttt{docker-compose.yml} file, and (b) a visual representation of the same file using \textit{Docker Composer}.}
  \label{fig:stack_comparison}
\end{figure*}

To more clearly compare and demonstrate the differences of representation between the conventional text-based approach and the designed visual approach, Fig.~\ref{fig:stack_comparison} shows a concrete example with a side-by-side comparison between the textual representation (a) and the equivalent visual representation (b). While it may not be immediately clear, both representations convey the same information. While the visual approach makes the artifacts themselves and their connections more evident, some properties (\textit{e.g.}, \texttt{stdin\_open} on the client service) will be shown when hovering some of the elements with a mouse pointer. The particular orchestration that is presented follows a client-server architecture comprised of three services: a web \textit{frontend} service (\textit{client}), a \textit{backend} web service (\textit{server}), and MongoDB database service (\textit{db}). We also include two custom networks, named \textit{private} and \textit{public}, to isolate the \textit{backend} from the \textit{frontend} as well as a named volume, called \textit{mongo-data}, for data persistence.

Docker Composer, and the entire low-code environment that it provides, differs from the tools that we review in Section~\ref{sec:related-work} in different aspects, and most notably in the support that it provides for the \textit{Docker Compose Specification}. The tools that are most closely-related to Docker Composer (\textit{DockStation}, \textit{Admiral} and \textit{Docker Studio}) do not offer a visual notation that covers all the elements that can be expressed in textual Docker Compose files, as we can see by the overview given in Table~\ref{tab:tools} (p.\,\pageref{tab:tools}).

\section{Empirical Study}
\label{sec:empirical-study}

With this study we seek to evaluate the viability and practical usefulness of a low-code environment in the domain of container orchestration. This is particularly relevant as empirical work in this field is still fairly limited~\cite{Rahman2019}. Although the theoretical benefits have been thoroughly evaluated in the past, there is still a severe lack of studies to assess whether these truly translate to practical scenarios.

The empirical study focuses on the evaluation of three activities in software engineering---analyzing, debugging, and implementing---in the context of Docker Compose configurations. A task was prepared for each activity, and we collected both performance-based and perception-based metrics. 

\subsection{Participants}

We selected a total of 16 students from the MSc in Informatics and Computing Engineering at the University of Porto, who volunteered to participate in the experiment. This methodology makes our sample a \textit{convenience sample}~\cite{baltes2020}. All participants had prior experience with Docker and Docker Compose due to their academic path and were randomly distributed between two groups, corresponding to the treatments: \textbf{control (CG)} and \textbf{experimental (EG)}. Both groups were asked to solve the same set of tasks. The participants of the \textbf{CG} had access to a text editor to edit the orchestration file and to a command-line shell to access the conventional toolchain. The participants of the \textbf{EG} had access to the experimental prototype to manage the orchestration as well as a command-line shell to execute additional commands if required (Docker related or not). In addition, both groups had complete access to the official Docker and Docker Compose documentation as well as any other resources on the internet.

\subsection{Environment}

The experimental sessions were conducted remotely. We opted for a remote workstation, set up in advance with the required software and materials, which was later made available to the participants. These resources included a browser (to access the experimental guidelines and surveys), a text editor set up in the appropriate directory, a command-line shell set up in the appropriate directory for both groups and the prototype tool for the \textbf{EG}.

\subsection{Task Definition}
\label{sec:task-definition}

As previously stated, the goal was to evaluate the behavior of the tool for three basic activities: analyzing, debugging, and implementing an orchestration file. This effort was translated into 4 tasks each featuring a corresponding scenario. In \textbf{Task 1 (T1)} a functioning Docker Compose configuration was provided and the goal was to analyze its structure and understand the overall behavior. In \textbf{Task 2 (T2)} a buggy configuration was provided and the goal was to debug and fix the faulty behavior. \textbf{Task 3 (T3)} focused on implementing and was divided into \textbf{T3.1}---build a simple configuration from the ground up, involving two containers, two environment variables, a custom network and a volume (implementation)---and \textbf{T3.2}---modify the configuration to use secrets instead of environment variables (increment).

\begin{figure}[b]
    \centering
    \includegraphics[width=0.9\linewidth]{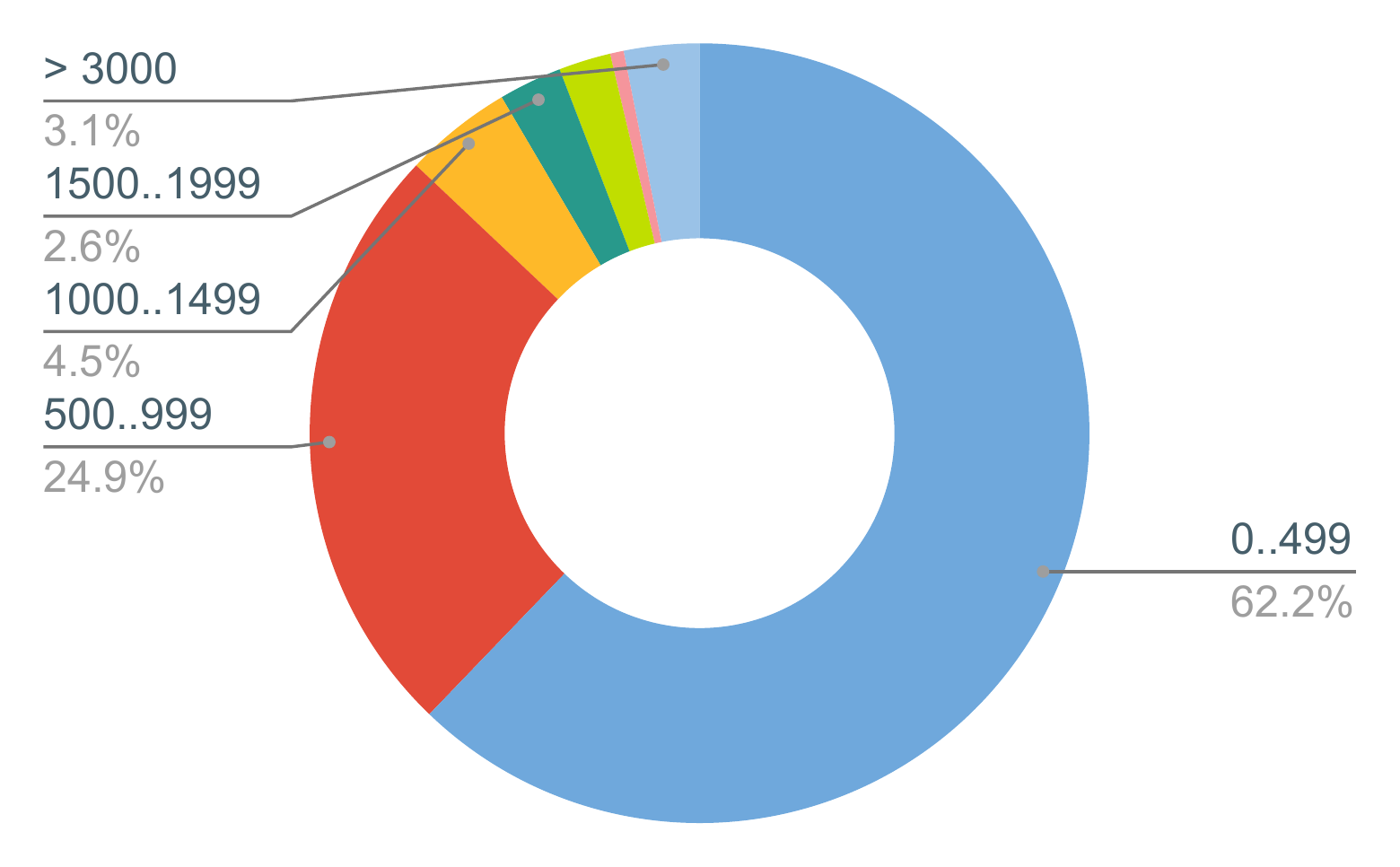}
    \caption{Distribution of 875 526 Docker Compose YAML files on Github by size, in bytes, as of January 2022.}
    \label{fig:size_overview}
\end{figure}

To ensure a balance between scale, complexity, realism, and expected time to completion in the tasks, we conducted a brief study to characterize the typical size of Docker Compose files. Namely, we tried to determine the typical number of containers in projects using Docker Compose, using GitHub as a data source. As depicted in Fig.~\ref{fig:size_overview}, approximately 62\% of the 875 526 considered files had sizes up to 500 bytes, which we equated to low complexity---typically containing one or two containers and minimal additional configurations. We gathered this data using the official code search API for files named specifically \texttt{docker-compose.yml} and, thus, limited by search mechanism both in terms of precision and number of results\footnote{GitHub Code Search, \url{https://docs.github.com/en/search-github/searching-on-github/searching-code}} .

\subsection{Procedure}
\label{sec:procedure}

A full session took between 50 minutes to 2 hours per participant. Each session was conducted individually with the researcher overseeing and observing the full procedure. Communication was done via remote voice call. The participants were encouraged to think aloud throughout the session so that the researcher could more clearly understand and follow along with their rationale. This strategy was also useful in identifying potentially unforeseen issues with the experiment's design.

Once the connection to the remote workstation was established, the participant had access to the instructions for the full procedure available in the remote environment. We make these instructions available as part of our replication package\footnote{\label{foot:package}A replication package to facilitate and encourage the independent replication of this experimental design is accessible at \url{https://doi.org/10.5281/zenodo.4001049}.}~\cite{bruno_piedade_2020_4001050}. The procedure was organized in the following steps:

\begin{itemize}
    \item \textbf{Background Survey.} This survey contained a set of questions to assess the current degree of experience with technologies which we had foreseen to potentially be confounding factors.
    
    \item \textbf{Tutorial.} Before solving the actual tasks, the participants had to follow a simple tutorial reviewing some basics of Docker Compose. This was mostly targeted to the \textbf{EG} so that they had some prior hands-on experience with the prototype. Nonetheless, to maintain consistency between both groups, participants in the \textbf{CG} also had to achieve the same goal with the conventional toolchain.
    
    \item \textbf{Experimental Tasks.} Participants were instructed to solve a set of four orchestration-related tasks. To maintain the total duration reasonable, time limits were set for each task. Participants were asked to advance to the next task whenever this time limit was exceeded.
    
    \item \textbf{Post-experiment Survey.} Participants were asked to fill a survey to assess their experience and evaluate the experience of working with the tools. The survey in the \textbf{EG} differed from the control since it included an additional set of questions to specifically evaluate the solution prototype.
\end{itemize}

\subsection{Research Variables}
\label{sec:variables}

We use both performance-based metrics and perception-based metrics as dependent variables in our study. 
    
The performance-based metrics consist of:

\begin{itemize}
    \item \textbf{Task Completion}, which refers to the ratio between the participants that successfully completed a task and the number of participants that tried to complete it.
    \item \textbf{Work Context Times} refers to the times spent on different work contexts, which we define later in this section.
    \item \textbf{Task Times} refers to the total time spent completing each task.
    \item \textbf{Execution Attempts} refers to the the number of times a participant tried to run the orchestration.
    \item \textbf{Context Switches} refers to the number of times participants accessed each of the contexts.
\end{itemize}

Some of these performance-based metrics lean on the different existing \textit{work contexts} that participants switched between when executing the tasks, and that we define as:
    \begin{itemize}
        \item \textbf{Script.} Time spent looking at the instructions and task description.
        \item \textbf{Documentation.} Time spent in the official Docker and Docker Compose documentation and Docker Hub.
        \item \textbf{Composer.} Time spent in the solution prototype, Docker Composer (only applicable to the experimental group)
        \item \textbf{Browser.} Time spent on the browser when accessing service's UIs and other documentation resources outside of those specified in the Documentation context.
        \item \textbf{Editor.} Time spent on the text editor to access and edit the materials.
        \item \textbf{Terminal.} Time spent on the terminal, mostly for executing Docker and Docker Compose CLI commands.
    \end{itemize}

The perception-based metrics we use were first introduced by Davis \textit{et al.}~\cite{davis1989perceived,davis1989user} and consist of:

\begin{itemize}
    \item \textbf{Perceived Ease of Use} (PEOU) refers to how much effort would be required to use the prototype.
    \item \textbf{Perceived Usefulness} (PU) refers to how well the prototype satisfies the participant's needs and expectations.
    \item \textbf{Intention to Use} (ITU) refers to the degree that the participant wishes to use the tool in the future.     
\end{itemize}

\subsection{Data Collection}

The results of the background questionnaire consist of answers of different types, including items using 5-point Likert scale, linear numeric scales, and multiple-choice questions.

Performance measurements for tasks were recorded manually by the researcher. An application named Turns Timer\footnote{Turns Timer, is an Android application available at \url{https://play.google.com/store/apps/details?id=com.deakishin.yourturntimer}} was used to register the time spent on individual activities, as well as the number of changes between contexts. This was achieved by attributing a timer for each context. The sum of all the timers was the total time spent on that task. 

Participants were asked to register the start and end time for each task in the form as a redundancy precaution in case some data was lost or incorrectly recorded by the researcher. In addition, the number of execution attempts was also registered by the researcher. These performance metrics, namely, durations and execution attempts, addressed \ref{hypothesis:RQ1} and \ref{hypothesis:RQ2}.

Participants were also asked to save their solutions in the workstation. This was done for subsequent review if needed. The solutions considered the answers given and the developed \texttt{docker-compose.yaml} files as requested in the tasks.

\ref{hypothesis:RQ3} was addressed through the post-experiment survey. This questionnaire mostly contained Likert-scale questions as well as a few open-ended questions. The former questions focused on the perception-based metrics---PEOU, PU and ITO (\textit{cf.} Section~\ref{sec:variables})--- for which we opted to follow a similar design to that employed by Sandobalin \textit{et al.}~\cite{Sandobalin2020}. 

It is important to note that we measure PEOU in both groups but measure PU and ITU exclusively in the experimental group. We adopt this approach for PU and ITU because these metrics intrinsically assume a subjective reference point. We believe that participants in the CG would state their perception in relative terms to the non-existence of Docker Compose, and participants in the EG would most likely state it in comparison to manipulating a \textit{docker-compose.yml} file directly. Participants could also share further observations and considerations in the open-ended questions. These were primarily useful in detecting potentially overlooked issues with the experimental procedure and even unforeseen validity threats.

\subsection{Pilot Experiments}
\label{sec:pilots}

We conducted two pilot experiments to gather feedback about the quality and consistency of the materials and of the experimental procedure itself. The first pilot allowed us to realise that some tasks were too complex to fit within the time of the experiment. As a result, we have redesigned and simplified them. The second pilot allowed to refine details in the materials, including typos and small inconsistencies, as well as to streamline the data collection process, in particular, the use of the Turns Timer application to register context times.

\section{Results and Discussion}
\label{sec:res}

The data collected was mainly quantitative, and we have used it for hypothesis testing, employing the \textit{Mann-Whitney~U} (MW-U)~\cite{mann1947test,wilcoxon1992individual} and \textit{McNemar}~\cite{mcnemar1947note} tests against our variables of interest. The notation used represents $H_0$ as the null hypothesis and $H_1$ as the alternative hypothesis, u for the U statistic of MW-U tests, and $\rho$ as the probability of rejecting $H_0$. We also denote $\sigma$ as the standard deviation and $\overline{x}$ as the mean.

\subsection{Background and Tutorial}
\label{sec:background}

The background survey gathers information about confounding factors to ensure that the groups are balanced in experience and skills. Questions are defined as Likert items and numeric values and inquire if the participants consider themselves experienced with (1) visual programming, (2) orchestration frameworks and tools, (3) Docker and (4) Docker Compose---configuration of volumes, networks, configs and secrets.

We show a summary of the results for the Likert and numeric scale questions in Table~\ref{tab:background}. Considering the alternative hypothesis stating that the control group is different from the experimental group (CG $\neq$ EG) for each of the background questions, we found \textbf{no significant difference} in experience or skills between the groups, except for BQ6. We discuss this difference at the end of this section.

\begin{table}[ht]
\centering
\addtolength{\tabcolsep}{-0.03cm}
\begin{tabular}{lrrrrrrcrr}
    & \multicolumn{2}{c}{CG}                        & \multicolumn{1}{l}{} & \multicolumn{2}{c}{EG}                        & \multicolumn{1}{l}{} & \multicolumn{3}{c}{MW-U}                          \\ \cline{2-3} \cline{5-6} \cline{8-10} 
    & \multicolumn{1}{c}{$\overline{x}$} & \multicolumn{1}{c}{$\sigma$} & \multicolumn{1}{l}{} & \multicolumn{1}{c}{$\overline{x}$} & \multicolumn{1}{c}{$\sigma$} & \multicolumn{1}{l}{} & $H_1$ & \multicolumn{1}{c}{u} & \multicolumn{1}{c}{$\rho$} \\ \hline
BQ1 & 2.88                  & 0.398                 &                      & 3.13                  & 0.398                 &                      & $\neq$ & 29.0                 & 0.372                 \\
BQ2 & 2.13                  & 0.581                 &                      & 1.63                  & 0.263                 &                      & $\neq$ & 30.5                 & 0.431                 \\
BQ3 & 4.00                  & 0.189                 &                      & 4.13                  & 0.350                 &                      & $\neq$ & 25.5                 & 0.214                 \\
BQ4 & 3.25                  & 0.412                 &                      & 3.13                  & 0.389                 &                      & $\neq$ & 30.5                 & 0.434                 \\
BQ5 & 3.25                  & 0.412                 &                      & 2.75                  & 0.458                 &                      & $\neq$ & 25.0                 & 0.223                 \\
BQ6 & 2.88                  & 0.295                 &                      & 1.63                  & 0.263                 &                      & $\neq$ & 9.0                  & 0.060                 \\
BQ7 & 4.38                  & 0.822                 &                      & 4.88                  & 0.515                 &                      & $\neq$ & 26.0                 & 0.262                 \\
BQ8 & 2.88                  & 0.895                 &                      & 2.63                  & 0.925                 &                      & $\neq$ & 31.0                 & 0.458                 \\
BQ9 & 3.50                  & 1.052                 &                      & 3.75                  & 0.675                 &                      & $\neq$ & 25.5                 & 0.244                 \\ \hline
\multicolumn{10}{l}{
	\scriptsize
	\begin{tabular}[c]{@{}l@{}}
	    I consider myself experienced with ...\\
		\textbf{BQ1.} ... visual programming tools.\\ 
		\textbf{BQ2.} ... with orchestration frameworks.\\ 
		\textbf{BQ3.} ... with the Linux OS.\\ 
		\textbf{BQ4.} ... with Docker.\\ 
		\textbf{BQ5.} ... with Docker Compose for development purposes.\\ 
		\textbf{BQ6.} ... with Docker Compose in production environments.\\ 
		Until now, approximately in how many projects have you ...\\
		\textbf{BQ7.} ... worked on which have used Docker Compose?\\ 
		\textbf{BQ8.} ... created/updated a docker-compose.yml file?\\ 
		\textbf{BQ9.} ... used docker-compose.yml files created by others?
	\end{tabular}
}
\\ \hline
\end{tabular}
\caption[Summary of the answers to background questionnaire]{Summary of the answers to the Likert and numeric scale questions in the background questionnaire.}
\label{tab:background}
\end{table}

The participants were also asked to specify what other orchestration frameworks, if any, had they used in the past. Only Kubernetes came up in the answers, with 2 participants of the CG and 3 of the EG reporting to have used it. Considering an alternative hypothesis that the control group is different from the experimental group (CG $\neq$ EG) for the number of participants that have used Kubernetes in the past, the results show \textbf{no a significant difference} of experience and skills between the groups (\textit{cf}. Table~\ref{tab:background_tools}). 

\begin{table}[H]
\centering
\addtolength{\tabcolsep}{-0.03cm}
\begin{tabular}{lrrrrrrcrr}
    & \multicolumn{2}{c}{CG}                        & \multicolumn{1}{l}{} & \multicolumn{2}{c}{EG}                        & \multicolumn{1}{l}{} & \multicolumn{3}{c}{MW-U}                          \\ \cline{2-3} \cline{5-6} \cline{8-10} 
    & \multicolumn{1}{c}{$\overline{x}$} & \multicolumn{1}{c}{$\sigma$} & \multicolumn{1}{l}{} & \multicolumn{1}{c}{$\overline{x}$} & \multicolumn{1}{c}{$\sigma$} & \multicolumn{1}{l}{} & $H_1$ & \multicolumn{1}{c}{u} & \multicolumn{1}{c}{$\rho$} \\ \hline
OF  & \multicolumn{1}{r}{0.25} & 0.463                 & \multicolumn{1}{r}{} & \multicolumn{1}{r}{0.38} & 0.518                 & \multicolumn{1}{r}{} & $\neq$ & 28                & 1.000                 \\ \hline
\multicolumn{9}{l}{\scriptsize\begin{tabular}[c]{@{}l@{}}\textbf{OF.} Number of orchestration frameworks used\end{tabular}}                      \\ \hline
\end{tabular}
\caption[Summary of the number of orchestration frameworks used]{Summary of the number of previously used tools specified in the background questionnaire.}
\label{tab:background_tools}
\end{table}

Another question inquired subjects about what individual Docker Compose configuration options they had configured in the past. The same number of participants in each group reported having configured \textit{secrets} and \textit{configs}. To confirm that there is not a significant difference in the use of \textit{volumes} and \textit{networks}, we ran a McNemar test. Considering an alternative hypothesis that the control group is different from the experimental group (CG $\neq$ EG), the results do \textbf{not show a significant difference} between the groups (\textit{cf}. Table~\ref{tab:background_artifacts_mcnemar}).

\begin{table}[H]
\centering
\begin{tabular}{lclclcr}
         & CG                       &                      & EG                       &                      & \multicolumn{2}{c}{McNemar} \\ \cline{2-2} \cline{4-4} \cline{6-7} 
         & \%                       &                      & \%                       &                      & $H_1$  & \multicolumn{1}{c}{$\rho$}  \\ \hline
Volumes  & \multicolumn{1}{r}{37.5} & \multicolumn{1}{r}{} & \multicolumn{1}{r}{62.5} & \multicolumn{1}{r}{} & $\neq$  & 0.687                  \\
Networks & \multicolumn{1}{r}{37.5} & \multicolumn{1}{r}{} & \multicolumn{1}{r}{25.0} & \multicolumn{1}{r}{} & $\neq$  & 1.000                  \\ \hline
\end{tabular}
\caption{Results of the McNemar test for configured Docker Compose options.}
\label{tab:background_artifacts_mcnemar}
\end{table}

To conclude the background analysis, taking into account all of the data collected and corresponding analysis, we believe that we can argue with some level of confidence that the subjects were balanced across both groups. Unfortunately, we cannot explain the answers to BQ6, and perhaps they translate a statistical anomaly, as they are not consistent with the answers pertaining to the number of projects (BQ7, BQ8, and BQ9), nor with the number of orchestration frameworks used (OF) or the Docker Compose options that participants have configured in the past.

To further ensure the groups were under equivalent conditions, before starting the experimental tasks, they have run a simple tutorial in the respective toolchain that they were requested to use.

\subsection{Experimental Tasks}

During the task we measured the \textit{task completion}, \textit{work context times}, \textit{task times}, \textit{execution attempts} and \textit{context switches}, as analysed next.

\subsubsection{Task Completion}
\label{sec:task-completion}

We have considered the effectiveness of task execution by looking at task completion. We define this metric as the ratio between successfully completed tasks and the total number of tasks. A task is successfully completed only if the subject finished within the allotted time limit and the solution was correct. 

\begin{figure}[h]
  \begin{center}
    \includegraphics[width=\linewidth]{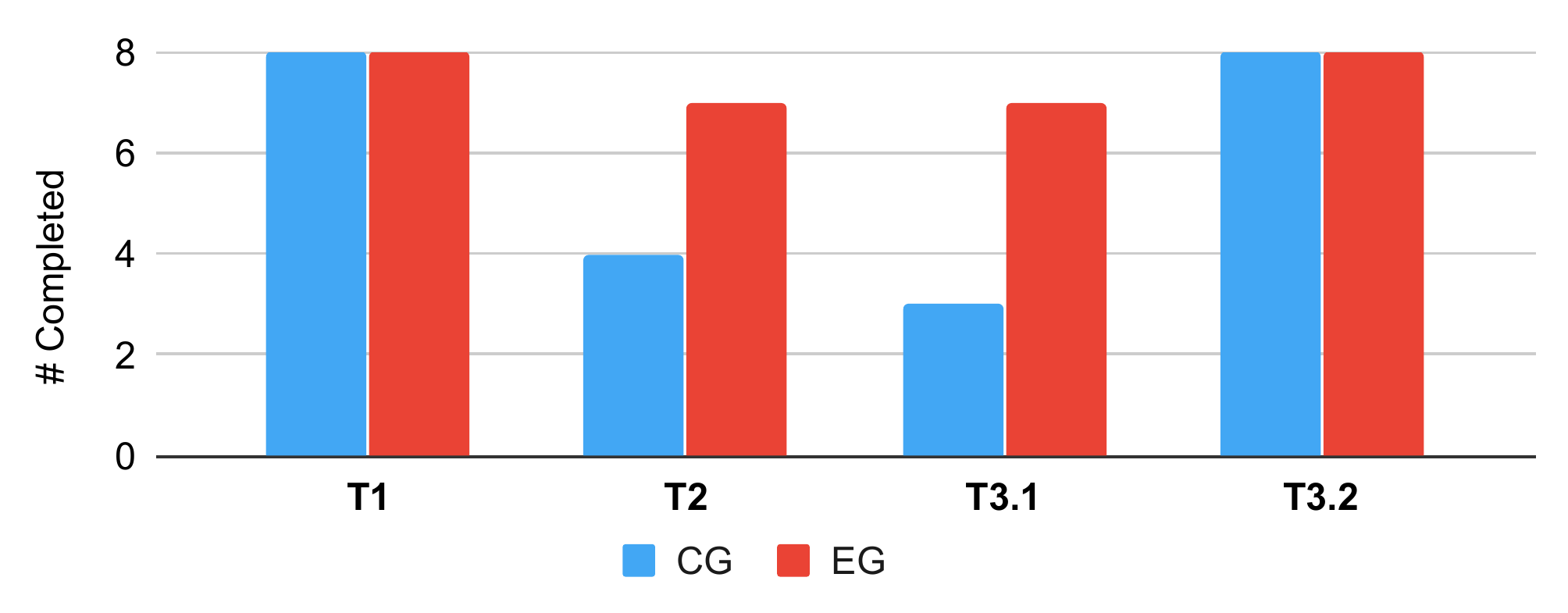}
    \caption{Distribution of total completed tasks per group.}
    \label{fig:task_effectiveness}
  \end{center}
\end{figure}

Fig.~\ref{fig:task_effectiveness} displays the distribution of completed tasks by group. While all participants completed T1 and T3.2, there is a clear difference in T2 and T3.1. While most of the participants in the \textbf{EG} completed the experimental tasks, only approximately half of the participants in the \textbf{CG} were able to complete them. The participants in both groups who were unable to complete the tasks were so due to the imposed time constraints on solving them. No case was registered in which the solution was incorrect. We can conclude that fewer participants in the \textbf{CG} finished the task T2 and T3.1. This, in turn, impacts the metrics considered for the remainder of this analysis since the registered times were capped up to the moment when the time limit was exceeded. If the time limit was not set, the differences might have been even sharper. However, this was a necessary sacrifice to keep the overall time reasonable and manageable.

\subsubsection{Work Context Times}
\label{sec:work-context-times}

The times spent on each work context allow us to understand the behavior of the participants better.

\begin{table*}[b]
\caption{Summary of the global time registered per activity for the sum of time taken in all tasks, with the mean and standard deviation for each group. The \textit{Composer} context does not contain data for the \textbf{CG} as this\\context was not available for this group and was exclusive to the \textbf{EG}.}
\label{tab:context_global}
\centering
\addtolength{\tabcolsep}{-0.01cm}
\begin{tabular}{lrrrrrrr}
                            & \multicolumn{3}{c}{CG}                                                & \multicolumn{1}{l}{} & \multicolumn{3}{c}{EG}                                                \\ \cline{2-4} \cline{6-8} 
\multicolumn{1}{c}{Context} & \multicolumn{1}{c}{$\sum$} & \multicolumn{1}{c}{$\overline{x}$} & \multicolumn{1}{c}{$\sigma$} & \multicolumn{1}{}{} & \multicolumn{1}{c}{$\sum$} & \multicolumn{1}{c}{$\overline{x}$} & \multicolumn{1}{c}{$\sigma$} \\ \hline
Script                      & 2:06:04               & 15:46               & 06:10               &                      & 1:29:40               & 11:13               & 03:52               \\
Composer   & \multicolumn{1}{c}{-}               & \multicolumn{1}{c}{-}               & \multicolumn{1}{c}{-}               &                      & 3:50:08               & 28:46               & 10:29               \\
Docs                        & 1:51:36               & 13:57               & 06:25               &                      & 0:18:59               & 02:22               & 02:15               \\
Browser                     & 0:41:44               & 05:13               & 04:02               &                      & 0:11:36               & 01:27               & 01:53               \\
Editor                      & 2:52:17               & 21:32               & 03:57               &                      & 0:04:51               & 00:36               & 00:29               \\
Terminal                    & 1:53:03               & 14:08               & 02:38               &                      & 0:02:10               & 00:16               & 00:31               \\
Stack Management            & 3:34:01               & 26:45              & 07:16               &                       & 3:54:59               & 29:22               & 10:47               \\ \hline
\end{tabular}
\end{table*}

\begin{figure*}[t]
  \begin{center}
    \leavevmode
    \includegraphics[width=0.85\textwidth]{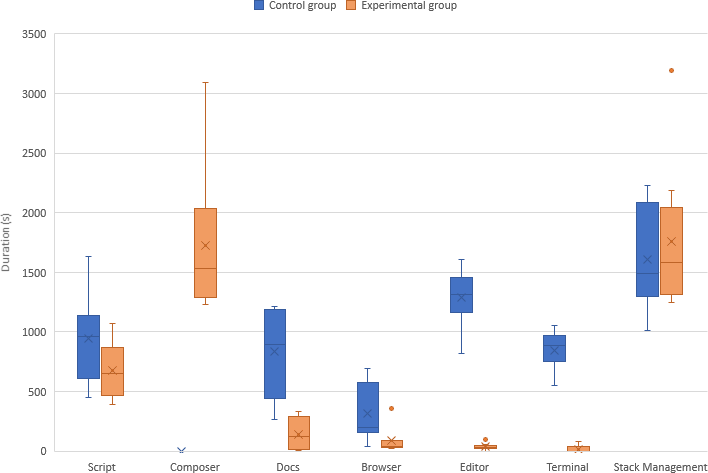}
    \caption[Distribution of the global times for each subject by context, by group]{Distribution of the global times for each subject by context, by group. The \textit{Stack Management} context refers to the sum of time spent on the \textit{Editor} and \textit{Terminal} contexts for the CG and the sum of time spent on the Editor and Composer contexts for the EG.}
    \label{fig:context_global_boxplot}
  \end{center}
\end{figure*}

Table~\ref{tab:context_global} and Fig.~\ref{fig:context_global_boxplot} overview the global times per context. We shall first look into those that are most directly comparable between the two groups---\textit{Script}, \textit{Docs} (documentation) and \textit{Browser}.

By looking at the data in Fig.~\ref{fig:context_global_boxplot}, we can identify a large discrepancy in the time spent on the \textit{Docs} context for reading documentation. This is further supported by the discrepancy of the time spent on the \textit{Browser} context, which was also mostly dedicated to reading other non-official documentation resources. We ran a Mann-Whitney U test for the independent contexts to confirm our intuition.

Considering the alternative hypothesis that the time spent in the \textit{Docs} and \textit{Browser} contexts is higher for the \textbf{CG}, the results shown in Table~\ref{tab:global_activities_mw} confirm that, indeed, the \textbf{CG} spent \textbf{significantly longer} than the \textbf{EG} in these contexts. We can also see that there is \textbf{not} a \textbf{significant difference} between the groups in the time spent on the \textit{Script} context for reading the script.

\begin{table}[h]
    \caption{Result of the Mann-Whitney U equality test for the sum of time spent on three contexts.}
    \label{tab:global_activities_mw}
    \centering
    \begin{tabular}{lcrr}
    \multicolumn{1}{c}{Context} & $H_1$              & \multicolumn{1}{c}{u} & \multicolumn{1}{c}{$\rho$} \\ \hline
    Script                      & \textgreater{} & 17                & 0.065                 \\
    Docs                        & \textgreater{} & 2                 & $<$0.001               \\
    Browser                     & \textgreater{} & 9                 & 0.007                 \\ \hline
    \end{tabular}
\end{table}%

It is difficult to draw any other useful information from the remainder of the variables when considered individually, as they are either exclusive to some group (\textit{i.e.}, \textit{Composer} for the EG) or partly replace the purpose of one another across both groups. However, we can consider the sum of time spent on \textit{editor} and \textit{terminal} (E+T) in the \textbf{CG} to be roughly equivalent to the sum of time spent on the textual \textit{editor} (which mostly equates to the time spent accessing other textual materials such as configuration files which were used in the tasks) and on \textit{Docker Composer} (E+C) in the \textbf{EG}. No participant in the \textbf{EG} used the terminal to execute any other Docker or Docker Compose CLI commands besides those that were available in the prototype. We refer to this composite context focused on the management of the orchestrations as \textit{Stack Management} and show it as the last line in Table~\ref{tab:context_global}. The time difference shown in this line does not appear to be very high. Testing the hypothesis (\textbf{CG $>$ EG} using the MW-U test rendered $u=29$ and $\rho=0.399$. These results do indeed \textbf{not show} that the participants in the \textbf{EG} have \textbf{spent significantly less time} managing the containers than those of the \textbf{CG}.

Therefore, it seems reasonable to conclude that the biggest impact on the overall duration was the time spent consuming documentation. This difference is in line with the expectation that a visual programming language promotes an exploratory approach in which the solution space is constrained by the options that are explicitly made available through the user interface, and users are able to converge to solutions by searching the options provided by our prototype.

\subsubsection{Task Times}
\label{sec:task-times}

Analyzing the times per context provides detailed insight into the participants' behavior. We can, however, also look at the time spent globally (\textit{i.e.}, the total sum of time spent on each activity) to assess the overall speed.

\begin{figure}[h]
  \begin{center}
    \leavevmode
    \includegraphics[width=\linewidth]{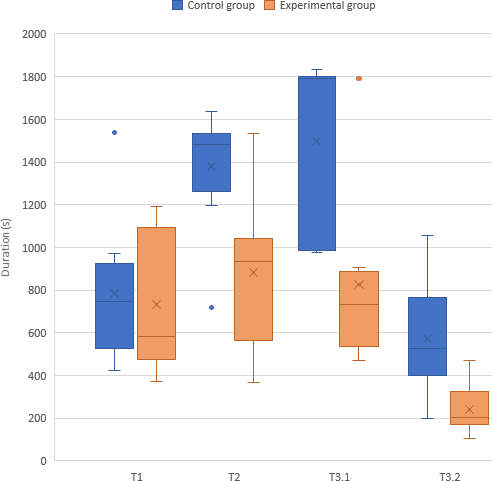}
    \caption{Distribution of times to completion for each subject by task, by group.}
    \label{fig:durations_boxplot}
  \end{center}
\end{figure}

Fig.~\ref{fig:durations_boxplot} displays the distribution of times by task for each group. We can identify that the participants in the \textbf{EG} generally have finished tasks T2, T3.1, and T3.2 sooner than the participants in the \textbf{CG}. In contrast, for task T1, both groups are more balanced. 

Table~\ref{tab:task_times} summarizes the results obtained for the times of each task along with the results of the MW-U significance test performed to compare both. By considering this data and the expected alternative hypothesis which states that the participants in the \textbf{EG} would finish tasks faster than those of the \textbf{CG} (\textit{i.e.}, CG $>$ EG) for all tasks, the results demonstrate that \textbf{EG} did indeed \textbf{finish task T2, T3.1 and T3.2 significantly faster} than the \textbf{CG}. These tasks evaluated debugging, implementing, and updating activities. Particularly, it is interesting to note the significant difference in T3.2. The scenario in this task required the participants to use a particular feature of Docker Compose---\textit{secrets}---with which most did in fact not have any prior experience. In practice, the workflow to use this feature in the prototype was very similar to that of other artifacts, such as volumes and networks. These results support that the prototype was sufficiently intuitive for participants to learn how to use this new feature, after having some experience with it, simply by following a similar rationale and without the need to consult additional documentation.

\begin{table}[h]
\caption[Summary of the times for each task across both groups]{Summary of the completion times for each task across groups.}
\label{tab:task_times}
\centering
\addtolength{\tabcolsep}{-0.03cm}
\resizebox{\linewidth}{!}{\begin{tabular}{lrrrrrrcrr}
                         & \multicolumn{2}{c}{CG}                        & \multicolumn{1}{l}{} & \multicolumn{2}{c}{EG}                        & \multicolumn{1}{l}{} & \multicolumn{3}{c}{MW-U}                                       \\ \cline{2-3} \cline{5-6} \cline{8-10} 
\multicolumn{1}{c}{Task} & \multicolumn{1}{c}{$\overline{x}$} & \multicolumn{1}{c}{$\sigma$} & \multicolumn{1}{l}{} & \multicolumn{1}{c}{$\overline{x}$} & \multicolumn{1}{c}{$\sigma$} & \multicolumn{1}{l}{} & $H_1$              & \multicolumn{1}{c}{u} & \multicolumn{1}{c}{$\rho$} \\ \hline
T1                       & 0:13:05               & 0:05:51               &                      & 0:12:12               & 0:05:19               &                      & \textgreater{} & 31                 & 0.480                 \\
T2                       & 0:22:59               & 0:04:55               &                      & 0:14:41               & 0:05:59               &                      & \textgreater{} & 11                 & 0.014                 \\
T3.1                     & 0:24:56               & 0:07:04               &                      & 0:13:47               & 0:06:57               &                      & \textgreater{} & 3                  & 0.001                 \\
T3.2                     & 0:09:35               & 0:04:26               &                      & 0:04:01               & 0:01:56               &                      & \textgreater{} & 6                  & 0.002                  \\ \hline
\end{tabular}}
\end{table}

The prototype successfully reduced the overall time required to develop and debug orchestrations. While there was no meaningful improvement for task T1 (in which participants had the goal of analyzing an orchestration), overall, the prototype managed to reduce the duration of the remaining tasks. Some of the questions in T1 required a deeper knowledge of concepts that were not immediately conveyed by the prototype. Although the participants in \textbf{EG} already had some hands-on experience with the prototype during the tutorial, we think that they spent some time exploring the features of the prototype, in search of answers for the first task.

\subsubsection{Execution Attempts}
\label{sec:execution-attempts}

In addition to the task times, the execution attempts were also registered for each task, that is, the number of times a participant tried to run the orchestration (\textit{i.e.}, run the command \texttt{docker-compose up}). 

\begin{figure}[h]
  \begin{center}
    \leavevmode
    \includegraphics[width=\linewidth]{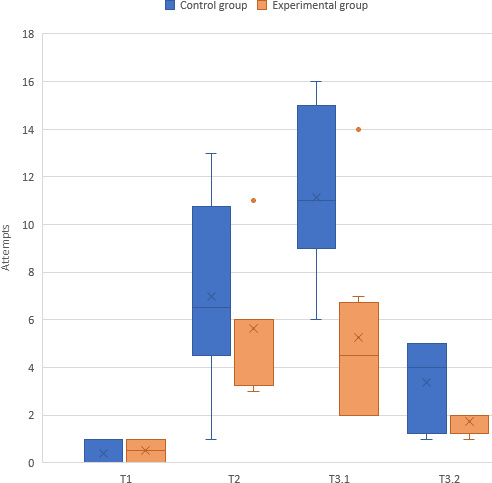}
    \caption{Distribution of execution attempts for each subject by task, by group.}
    \label{fig:execs_dist}
  \end{center}
\end{figure}

Fig.~\ref{fig:execs_dist} displays the distribution of execution attempts by task for each group. We can see that the participants in the \textbf{EG} have generally performed fewer execution attempts in tasks T2, T3.1, and T3.2 than the participants in the \textbf{CG}. In contrast, for task T1, both groups are more balanced, but it is important to note that the results for T1 are not very revealing as the execution of the orchestration was completely optional for this task.

Table~\ref{tab:task_execs} displays the results for execution attempts. Considering this data and the expected alternative hypothesis which states that the \textbf{EG} would need fewer execution attempts than the \textbf{CG} (\textit{i.e.}, \textbf{CG} $>$ EG) for all tasks, the results demonstrate that \textbf{EG} did require \textbf{significantly fewer execution attempts for T3.1 and T3.2}.

\begin{table}[h]
\caption[Summary of the execution attempts for each task across both groups]{Summary of the execution attempts for each task across both groups.}
\label{tab:task_execs}
\centering
\resizebox{\linewidth}{!}{\begin{tabular}{lrrrrrrcrr}
                         & \multicolumn{2}{c}{CG}                        & \multicolumn{1}{l}{} & \multicolumn{2}{c}{EG}                        & \multicolumn{1}{l}{} & \multicolumn{3}{c}{MW-U}                                       \\ \cline{2-3} \cline{5-6} \cline{8-10} 
\multicolumn{1}{c}{Task} & \multicolumn{1}{c}{$\overline{x}$} & \multicolumn{1}{c}{$\sigma$} & \multicolumn{1}{l}{} & \multicolumn{1}{c}{$\overline{x}$} & \multicolumn{1}{c}{$\sigma$} & \multicolumn{1}{l}{} & $H_1$              & \multicolumn{1}{c}{u} & \multicolumn{1}{c}{$\rho$} \\ \hline
T1                       & 0.38                  & 0.518                 &                      & 0.50                  & 0.535                 &                      & \textgreater{} & 28.0                 & 0.500                 \\
T2                       & 7.00                  & 3.928                 &                      & 5.63                  & 2.560                 &                      & \textgreater{} & 21.5                 & 0.134                 \\
T3.1                     & 10.13                 & 4.357                 &                      & 5.25                  & 4.097                 &                      & \textgreater{} & 11.5                 & 0.014                 \\
T3.2                     & 3:50                  & 1.690                 &                      & 1.75                  & 0.463                 &                      & \textgreater{} & 13.0                 & 0.016                  \\ \hline                         
\end{tabular}}
\end{table}

These results are in line with the time difference established above. Overall, the participants in the \textbf{EG} were more efficient and did not spend as much time restarting the containers. This behavior was also expected as in practice, many execution attempts in the \textbf{CG} resulted from syntax errors. The prototype avoided most syntax errors simply due to the more strict form inputs (with stronger validation) and subsequent automatic code generation, free of errors. We believe that this was the biggest factor contributing to the non-significant difference in T2 since a partially working orchestration was provided in this task, and it required few changes.

\subsubsection{Context Switches}
\label{sec:context-switches}

In addition to the work context times, the \textbf{context switches} were also recorded, that is, the number of times the participant accessed each of the contexts. To keep this metric uniform across participants, we consider the context switches per minute (s/m) instead of to the total count of context switches. This metric is useful in evaluating participants' degree of focus when using the tool. We argue that a higher number of context switches translates into a less optimized experience since users have to shift their attention more frequently.

\begin{figure}[h]
  \begin{center}
    \leavevmode
    \includegraphics[width=0.75\linewidth]{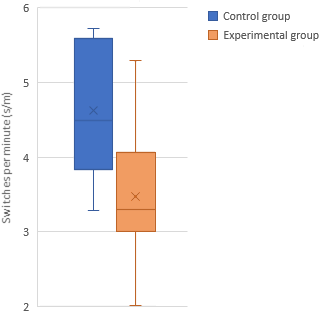}
    \caption{Distribution of global context switches for each subject by group.}
    \label{fig:switches_global}
  \end{center}
\end{figure}

We analyze the global context switches during the full session, that is, the total sum of the switches between all contexts for all tasks. Fig.~\ref{fig:switches_global} seems to suggest that the participants in the \textbf{EG} performed fewer context switches than those of the \textbf{CG}. To confirm this intuition, we performed a MW-U test (\textit{cf}. Table~\ref{tab:switches_global_mw}). Considering this data and the alternative hypothesis which states that participants in the \textbf{EG} would execute fewer context switches than those in the \textbf{CG} (\textit{i.e.}, \textbf{CG} $>$ \textbf{EG}) overall, the results show that the participants in the \textbf{EG} did, in fact, \textbf{execute significantly fewer context switches} than those in the \textbf{CG}. These results suggest that the process was more streamlined for the \textbf{EG}, which is in line with the results of the task time analysis performed previously. 

Notwithstanding, to interpret these results, we must also consider that CG participants used external terminal windows to execute CLI commands, rather than a built-in terminal within the text editor. It is reasonable to expect that we would see a smaller difference in context switching across the two groups if we were to condition the CG to use a built-in terminal.

\begin{table}[h]
\caption{Results of the MW-U test for global context changes.}
\label{tab:switches_global_mw}
\centering
\resizebox{\linewidth}{!}{\begin{tabular}{lcclcclccc}
                        & \multicolumn{2}{c}{CG}                                &                      & \multicolumn{2}{c}{EG}                                &                      & \multicolumn{3}{c}{MW-U}                                         \\ \cline{2-3} \cline{5-6} \cline{8-10} 
\multicolumn{1}{c}{s/m} & $\overline{x}$                         & $\sigma$                         &                      & $\overline{x}$                         & $\sigma$                         &                      & $H_1$           & u                      & $\rho$                         \\ \hline
Global                  & \multicolumn{1}{r}{4.628} & \multicolumn{1}{r}{0.912} & \multicolumn{1}{r}{} & \multicolumn{1}{r}{3.479} & \multicolumn{1}{r}{0.929} & \multicolumn{1}{r}{} & \textless{} & \multicolumn{1}{r}{10} & \multicolumn{1}{r}{0.010} \\ \hline
\end{tabular}}
\end{table}

\subsection{Assessment Survey}
\label{sec:assessment-survey}

This survey had the goal of gathering insights on the perception of participants regarding aspects of the experiment itself and of the visual approach that we aimed to evaluate. We start by asking how participants perceived the environment (\textit{cf}.~Section~\ref{sec:survey-environment}) and the clarity of the instructions and task descriptions (\textit{cf}.~Section~\ref{sec:survey-instructions}), with the goal of detecting unforeseen confounding factors. The subsequent questions intend to collect data about three perception-based metrics: perceived ease of use (PEOU, \textit{cf}.~Section~\ref{sec:survey-peou}), perceived usefulness (PU, \textit{cf}.~Section~\ref{sec:survey-pu}), and intention to use (ITU, \textit{cf}.~Section~\ref{sec:survey-itu}). The survey provided to \textbf{EG} featured an additional set of questions to evaluate the perceived usefulness of individual features and overall PU as well as ITU sentiment towards the prototype. The questions which focused on PU were formulated to compare the usefulness of the prototype in relation to the participant's perception of the conventional method and toolchain.

\subsubsection{Environment}
\label{sec:survey-environment}
Analyzing the data in Table~\ref{tab:ass_env} and considering an alternative hypothesis that the perception of the \textbf{CG} of environment factors differs from the \textbf{EG} for all environment-related questions, the results demonstrate that there is \textbf{not a significant difference} between the groups. These results support the hypothesis that the influence of environmental factors on performance during tasks was balanced across both groups and therefore, did not have a meaningful impact on the outcomes.

\begin{table}[h]
\caption{Summary of the answers to the ENV Likert-scale items of the assessment survey.}
\label{tab:ass_env}
\centering
\resizebox{\linewidth}{!}{\begin{tabular}{lcrlcrlcrr}
     & \multicolumn{2}{c}{CG}                           &                      & \multicolumn{2}{c}{EG}                           &                      & \multicolumn{3}{c}{MW-U}                          \\ \cline{2-3} \cline{5-6} \cline{8-10}
     & $\overline{x}$                        & \multicolumn{1}{c}{$\sigma$} &                      & $\overline{x}$                        & \multicolumn{1}{c}{$\sigma$} &                      & $H_1$ & \multicolumn{1}{c}{u} & \multicolumn{1}{c}{$\rho$} \\ \hline 
ENV1 & \multicolumn{1}{r}{4.13} & 1.356                 & \multicolumn{1}{r}{} & \multicolumn{1}{r}{3.75} & 1.282                 & \multicolumn{1}{r}{} & $\neq$ & 25.5                 & 0.231                 \\
ENV2 & \multicolumn{1}{r}{1.88} & 1.356                 & \multicolumn{1}{r}{} & \multicolumn{1}{r}{2.13} & 1.458                 & \multicolumn{1}{r}{} & $\neq$ & 29                 & 0.367                 \\ \hline
\multicolumn{10}{l}{\scriptsize\begin{tabular}[c]{@{}l@{}}\textbf{ENV1.} It was easy working in the remote machine.\\ \textbf{ENV2.} The environment was distracting.\end{tabular}}                                                                      \\ \hline
\end{tabular}}
\end{table}

\subsubsection{Clarity of the Instructions and Task Descriptions}
\label{sec:survey-instructions}

We found a \textbf{significant difference} regarding how the task descriptions were perceived across both groups, as shown by Table~\ref{tab:ass_ppu}, and the means suggest that the \textbf{EG} found the task descriptions more understandable than the \textbf{CG}. This discrepancy implies a relevant threat to validity, since it could entail that differences in performance reflect an intrinsic difficulty by the CG in understanding the instructions. However, the balance between the two groups that we report in Section~\ref{sec:background} makes us believe that that is unlikely. Another explanation that we must consider is that this difference in perception is the result of the more pronounced difficulties of the \textbf{CG} in executing the task---that is, in moving from the \textit{problem space} to the \textit{solution space}---and that this difficulty may have influenced their judgment about the instructions. The prototype used by the \textbf{EG} provided a more streamlined and focused experience (as supported by the lower context switching) which, we think, has helped participants to concentrate on the provided instructions and take a more linear approach to performing the tasks.

\begin{table}[h]
\caption{Summary of the answers to the CLR Likert-scale items of the assessment survey.}
\label{tab:ass_ppu}
\centering
\resizebox{\linewidth}{!}{\begin{tabular}{lcrlcrlcrr}
     & \multicolumn{2}{c}{CG}                           &                      & \multicolumn{2}{c}{EG}                           &                      & \multicolumn{3}{c}{MW-U}                          \\ \cline{2-3} \cline{5-6} \cline{8-10}
     & $\overline{x}$                        & \multicolumn{1}{c}{$\sigma$} &                      & $\overline{x}$                        & \multicolumn{1}{c}{$\sigma$} &                      & $H_1$ & \multicolumn{1}{c}{u} & \multicolumn{1}{c}{$\rho$} \\ \hline 
CLR1 & \multicolumn{1}{r}{2.25} & 1.282                 & \multicolumn{1}{r}{} & \multicolumn{1}{r}{1.25} & 0.463                 & \multicolumn{1}{r}{} & $\neq$ & 17.0                 & 0.080                 \\
CLR2 & \multicolumn{1}{r}{3.00} & 0.926                 & \multicolumn{1}{r}{} & \multicolumn{1}{r}{1.50} & 0.756                 & \multicolumn{1}{r}{} & $\neq$ & 6.5                  & 0.005                 \\ \hline
\multicolumn{10}{p{\linewidth}}{\scriptsize\begin{tabular}[c]{@{}p{\linewidth}@{}} \textbf{CLR1.} I found the procedure instructions complex and hard to follow\\ \textbf{CLR2.} I found the task descriptions complex and hard to follow.\end{tabular}}                                                                     \\ \hline
\end{tabular}}
\end{table}

\subsubsection{Perceived Ease of Use}
\label{sec:survey-peou}

By analyzing the data in Table~\ref{tab:ass_peou} and considering the hypothesis that participants in the \textbf{EG} would find the prototype easier to use (\textit{i.e.}, \textbf{CG} $>$ \textbf{EG} for PEOU1 and PEOU2 and \textbf{CG} $<$ \textbf{EG} for PEOU3) for all equivalent PEOU questions, the results demonstrate that the \textbf{EG} did indeed find that it was \textbf{significantly easier to work with the prototype}. Therefore we can state with some confidence that participants did find the prototype easier to use than the conventional method. Additionally, the exclusive question PEOU4 also demonstrates that the participants in \textbf{EG} strongly agreed that the tool is easy to learn.

\begin{table}[h]
\caption{Summary of the results of the answers to the Likert-scale questions related to perceived ease of use (PEOU) in the assessment survey. (*) PEOU4 was exclusive for the EG.}
\label{tab:ass_peou}
\centering
\resizebox{\linewidth}{!}{\begin{tabular}{lcrlcrlcrr}
     & \multicolumn{2}{c}{CG}                           &                      & \multicolumn{2}{c}{EG}                           &                      & \multicolumn{3}{c}{MW-U}                          \\ \cline{2-3} \cline{5-6} \cline{8-10}
     & $\overline{x}$                        & \multicolumn{1}{c}{$\sigma$} &                      & $\overline{x}$                        & \multicolumn{1}{c}{$\sigma$} &                      & $H_1$ & \multicolumn{1}{c}{u} & \multicolumn{1}{c}{$\rho$} \\ \hline 
PEOU1  & \multicolumn{1}{r}{2.64} & 1.188                 &                      & 1.00                  & 0.000                 & \multicolumn{1}{r}{} & \textgreater{} & 8.0                  & 0.002                 \\
PEOU2  & \multicolumn{1}{r}{2.63} & 0.916                 &                      & 1.13                  & 0.354                 & \multicolumn{1}{r}{} & \textgreater{} & 5.5                  & 0.001                 \\
PEOU3  & \multicolumn{1}{r}{3.63} & 0.744                 &                      & 5.00                  & 0.000                 &                      & \textless{} & 4.0                  & 0.001                 \\
PEOU4* & n/a                        & \multicolumn{1}{c}{n/a} &                      & 4.88                  & 0.354                 &                      & n/a              & \multicolumn{1}{c}{n/a} & \multicolumn{1}{c}{n/a} \\ \hline
\multicolumn{10}{l}{\scriptsize\begin{tabular}[c]{@{}l@{}} \textbf{PEOU1.} Overall, I found the tool difficult to use.\\ \textbf{PEOU2.} I found it difficult to understand stacks with the tool.\\ \textbf{PEOU3.} I found it easy to define stacks with the tool.\\ \textbf{PEOU4.} Overall, I found the tool easy to learn.\end{tabular}}                                                                                 \\ \hline
\end{tabular}}
\end{table}

\subsubsection{Perceived Usefulness}
\label{sec:survey-pu}

This and the next section are about questions exclusive to the survey provided to the EG, as they were specifically about Docker Composer, and could only be answered by the group that used it.

The survey provided to the \textbf{EG} contained a section dedicated to evaluating the perceived usefulness of our prototype when compared with a usual workflow without Docker Composer. The questions focused on specific features, namely, the visual map of artifacts (VM), Docker Hub integration (DHI), visual feedback (VF), and executing commands on the UI (UIC). They are used to assess the perceived usefulness with more granularity and support us in better understanding the impact of each feature in the overall perception. 

Table~\ref{tab:ass_features} summarizes the obtained results and allows us to conclude that the feature considered most useful was the visual map of artifacts (VM) while the least was the Docker Hub integration (DHI). These results match our expectations as the DHI feature was secondary and mostly added for ease-of-use and convenience. The designed tasks did not take full advantage of this feature since participants could copy and paste the image names and tags from the provided script without the need to locate them manually. In contrast, the VM feature was the direct result of the hypothesis of this dissertation and corresponded to the most novel and premeditated feature. Regardless, the response was positive for all features.

\begin{table}[h]
\caption[Summary of the answers to the feature items of the assessment survey]{Summary of the results of the answers to the Likert-scale questions related to perceived usefulness of features in the assessment survey. n/a means \textit{not applicable}.}
\label{tab:ass_features}
\addtolength{\tabcolsep}{-0.0085cm}
\centering
\resizebox{\linewidth}{!}{\begin{tabular}{lrrrrrrcclcc}
    & \multicolumn{2}{c}{VM}                        & \multicolumn{1}{l}{} & \multicolumn{2}{c}{DHI}                       & \multicolumn{1}{l}{} & \multicolumn{2}{c}{VF}                               &  & \multicolumn{2}{c}{UIC}                              \\ \cline{2-3} \cline{5-6} \cline{8-9} \cline{11-12} 
    & \multicolumn{1}{c}{$\overline{x}$} & \multicolumn{1}{c}{$\sigma$} & \multicolumn{1}{l}{} & \multicolumn{1}{c}{$\overline{x}$} & \multicolumn{1}{c}{$\sigma$} & \multicolumn{1}{l}{} & $\overline{x}$                        & $\sigma$                         &  & $\overline{x}$                        & $\sigma$                         \\ \hline
ULE & 4.88                  & 0.35                 &                      & 4                  & 1.20                 &                      & \multicolumn{1}{r}{4.50} & \multicolumn{1}{r}{0.76} &  & n/a                        & n/a                         \\
UQ  & 4.75                  & 0.46                 &                      & 4                  & 1.20                 &                      & \multicolumn{1}{r}{4.75} & \multicolumn{1}{r}{0.71} &  & n/a                        & n/a                        \\
DLE & 4.88                  & 0.35                 &                      & 4                  & 1.20                 &                      & n/a                        & n/a                         &  & \multicolumn{1}{r}{4.75} & \multicolumn{1}{r}{0.71} \\
DQ  & 4.88                  & 0.35                 &                      & 4                  & 1.20                 &                      & n/a                        & n/a                         &  & \multicolumn{1}{r}{4.75} & \multicolumn{1}{r}{0.71} \\ \hline
\multicolumn{12}{l}{\scriptsize\begin{tabular}[c]{@{}l@{}}
I find the [VM|DHI|VF|UIC] ...\\
\textbf{ULE.} ... helpful to understand stacks with less effort.\\ 
\textbf{UQ.} ... helpful to understand stacks more quickly.\\ 
\textbf{DLE.} ... helpful to define stacks with less effort.\\ 
\textbf{DQ.} ... helpful to define stacks more quickly.\end{tabular}}                                              \\ \hline
\end{tabular}}
\end{table}

The answers to questions about the overall perceived usefulness are shown in Table~\ref{tab:ass_pu}. Taking into account this feedback and that about the usefulness of individual features (\textit{cf}.~Table~\ref{tab:ass_features}), we can state that participants did indeed find the tool useful.

\begin{table}[h]
\caption[Summary of the answers to the PU items of the assessment survey]{Summary of the results of the answers to the Likert-scale questions related to the overall perceived usefulness in the assessment survey.}
\label{tab:ass_pu}
\centering
\resizebox{\linewidth}{!}{\begin{tabular}{p{2cm} l c c p{2cm}}
 &     & \multicolumn{1}{c}{$\overline{x}$} & \multicolumn{1}{c}{$\sigma$} & \\ \cline{2-4}
 & PU1 & 5.00 & 0.000 & \\
 & PU2 & 5.00 & 0.000 & \\
 & PU3 & 1.50 & 0.756 & \\
 & PU4 & 1.00 & 0.000 & \\
 & PU5 & 4.88 & 0.354 & \\ \hline
\multicolumn{5}{p{\linewidth}}{\scriptsize\begin{tabular}[c]{@{}p{\linewidth}@{}} 
\textbf{PU1.} I believe this tool would reduce the effort required to define stacks.\\ \textbf{PU2.} Overall, I found the tool useful.\\\textbf{PU3.} A stack visualized with the tool would be more difficult to understand. \\ \textbf{PU4.} Overall, I think this tool is ineffective for defining stacks.\\\textbf{PU5.} Overall, I think this tool improves the stack definition process.\end{tabular}}                                                                     \\ \hline
\end{tabular}}
\end{table}

\subsubsection{Intention to Use}
\label{sec:survey-itu}

Taking into account the results displayed in Table~\ref{tab:ass_itu}, we can state with some confidence that participants do indeed intend to use the tool in the future.

\begin{table}[h]
\caption[Summary of the answers to the ITU items of the assessment survey]{Summary of the results of the answers to the Likert-scale questions related to the overall perceived usefulness in the assessment survey.}
\label{tab:ass_itu}
\centering
\resizebox{\linewidth}{!}{\begin{tabular}{p{2cm} l r r p{2cm}}
 &      & \multicolumn{1}{c}{$\overline{x}$} & \multicolumn{1}{c}{$\sigma$} & \\ \cline{2-4}
 & ITU1 & 4.50 & 0.535 &  \\
 & ITU2 & 4.75 & 0.463 &  \\
 & ITU3 & 5.00 & 0.000 &  \\
 & ITU4 & 4.50 & 0.756 &  \\
 & ITU5 & 4.75 & 0.463 &  \\ \hline
\multicolumn{5}{p{\linewidth}}{\scriptsize\begin{tabular}[c]{@{}p{\linewidth}@{}} \textbf{ITU1.} This tool would make it easier for practitioners to define orchestrations.\\ \textbf{ITU2.} Using this tool would make it easier to explain the stack.\\\textbf{ITU3.} I would recommend this tool to work with Docker Compose.\\ \textbf{ITU4.} I would like to use this tool in the future.\\\textbf{ITU5.} It would be easy for me to become skillful in using this tool.\end{tabular}}                                                                     \\ \hline
\end{tabular}}

\end{table}

Overall, the results demonstrate that the response to the prototype was overwhelmingly positive and generally very consistent across participants. The participants found the approach more straightforward to use than the conventional method, generally useful, and were interested in using it in the future.

\section{Validation Threats}
\label{sec:validationthreats}

We identify and discuss threats that might hinder the soundness of the obtained results.
We tried to mitigate most threats throughout the experimental planning and design. Nevertheless, we identify the following threats.

\subsection{Internal Validity}

\noindent \textbf{Psychological bias.} For the results to be unbiased, it is important to ensure that participants are unaware of what group they belong to. However, this is hard to achieve in practice. Participants may have suspected they were part of the EG, since they were asked to use a tool that was not known to them. Nevertheless, all efforts were made to mitigate this threat, particularly by preparing the materials to omit any relevant information and avoiding any verbal exchange during the experiments themselves, which could allude to this fact.
    
\noindent \textbf{Experience differences.} It is crucial to ensure that the results are independent of possible skills and experience differences between groups. For this reason, the background questionnaire was part of the process, and the data analysis supports that both groups were balanced. Therefore, we believe we can discard this threat.

\noindent \textbf{Environment influences.} Performing the sessions remotely raised additional concerns in regards to possible deviations due to uncontrolled external factors. However, we believe that we were able to ensure a consistent environment for all participants with this approach. In addition, the researcher's constant observation during the sessions was also useful in identifying any unforeseen anomalies. The results in the assessment questionnaire further support that there was not any significant difference between groups. Thus, we believe we can discard this threat.

\subsection{External Validity}

\noindent \textbf{Limited sample size.} The somewhat small sample size limits the extent to which we can confidently assert the generalizability of our findings. To do so, replications of these studies should try to use a larger sample size to mitigate this threat.

\noindent \textbf{Sample bias.} All the participants were students with similar backgrounds. While this helped to ensure that there was not a significant experience disparity between groups, the results may be biased towards novice software developers. Research conducted by Host \textit{et al.}~\cite{Host2000} concluded that graduate students can be appropriate subjects, if properly trained, as they represent the future generations of developers. Salman \textit{et al.}~\cite{salman2015students} also conclude that, independently of the experience level, subjects perform similarly when they apply a new approach for the first time. Thus, we believe the results are meaningful. Despite this, further studies may obviously provide more insights, especially if they feature a more heterogeneous sample, with more diverse backgrounds, to achieve results with more confidence and mitigate this threat.

\subsection{Construct Validity}

\noindent \textbf{Clarity of the questions.} There is always the chance that some participants have interpreted our questionnaire items in a different way than the one we intended. To address this concern, we have first run two pilot experiments, as described in Section~\ref{sec:pilots}. This allowed improving potentially dubious questions before running the experiment.

\noindent \textbf{Auto-layout inefficiencies.} As described in Section~\ref{sec:dockercomposer}, the prototype implemented an automatic layout algorithm to position artifacts when loading an orchestration from a YAML file. However, the layouts achieved were not as good as if manually constructed. To mitigate possible deviations resulting from this limitation, the orchestrations provided to the experimental group were prepared in advance to a more readable format and were loaded using the custom storage feature. In practice, this did not seem to influence the results as no participant suggested this improvement, therefore we believe that we can discard this threat.

\section{Conclusions}
\label{sec:conclusions}

Our findings support some of the benefits that we expected to have from using low-code in this domain and, in particular, support the hypothesis that a visual approach for orchestration can indeed reduce development time and error-proneness significantly. We delve into this hypothesis in more detail by answering the three research questions that we first introduce in Section~\ref{sec:research-goals}.

\begin{enumerate}[label=\textbf{RQ\arabic*},leftmargin=2\parindent]
    \item \textit{To what extent does a visual notation for the orchestration of (Docker) containers reduce the development time?}
    
    \textbf{Answer:} The \textit{work context times} show a positive impact of Docker Composer, but the clearest gains are specifically in the reduction of the time spent reading documentation (\textit{cf}.~Section~\ref{sec:work-context-times}). Nevertheless, a clear improvement can be observed in \textit{task times}, and the most visible benefits appear specifically in tasks involving the development or debugging of orchestrations (\textit{cf}.~Section~\ref{sec:task-times}).

    \item \textit{To what extent does a visual notation for (Docker) orchestration files reduce the number of errors?}
    
    \textbf{Answer:} Both the analysis of the \textit{execution attempts} (\textit{cf}. Section~\ref{sec:execution-attempts}) and of the \textit{context switches} (\textit{cf}. Section~\ref{sec:context-switches}) suggests that the experience was overall more streamlined when using a visual notation. Docker Composer,  allowed to spend less time restarting containers and to avoid syntax errors, which resonates with the significantly lower number of context switches.
    
    \item \textit{What is the perception of developers towards using a visual notation for the orchestration of (Docker) orchestration files?}
    
    \textbf{Answer:} The findings related to perception-based metrics were positive overall (\textit{cf}. Section~\ref{sec:assessment-survey}). The participants felt that Docker Composer was more comfortable to use, was generally useful, and they showed strong intentions of using it in the future. These results give us some confidence that developers find the tool easy-to-use and intuitive, considering the steps and actions needed to configure some orchestrations successfully. 
\end{enumerate}

\subsection{Future Work}

While the results that we have obtained are promising, we do not consider the prototype production-ready. Additional research would be useful to consolidate further and confirm our findings, primarily to address some of the validation threats described above. To this end, we make our prototype and experimental package readily available and provide a roadmap along three different topics---the visual approach, the developed prototype, and the empirical evaluation.

\subsubsection{Visual Approach}

The designed visual approach is highly tied to the underlying concepts of Docker Compose, but we believe that similar visual approaches are applicable to a broader context and, in particular, to other orchestration technologies. Therefore, it would be interesting to explore domain-specific visual notations for other orchestration technologies (\textit{e.g.}, Kubernetes). Furthermore, one can even consider the possibility of a more generic and technology agnostic model-driven approach, useful for a wider set of use-cases. In conjunction with the growing adoption of microservices architectures, the positive results obtained in this work provide strong motivation to promote research in this field.

\subsubsection{Prototype}

We propose evolving the prototype to a production-ready application by refining and expanding existing features and exploring other ideas that go beyond the immediate objective of this work, which we believe may further improve the orchestration process. Some of these ideas stem from the conceptual stages of our implementation but were ultimately not realized as they did not directly contribute towards our goal, while others result from how we foresee the prototype could evolve.

\begin{itemize}
    \item \textbf{Textual editor.} We propose the inclusion of a textual editor, which would work in parallel with the graphical editor, as the similar feature offered by DockStation. However, unlike DockStation, we propose to view both perspectives on the same window simultaneously. This would require real-time sync mechanisms to maintain both views consistent on change in either one and could be achieved through techniques for MDSE bidirectional transformation~\cite{Stevens2007,Angyal2008,Hidaka2011,Hoisl2015} and flexible modeling~\cite{correia2013}. We consider this as one of the most significant improvements since it could broaden the potential target audience of the prototype, as many developers (especially the more experienced ones) seem to prefer working with text files. This addition would mean that the solution would not substitute the conventional method and instead complement it with more information and options.  
    \item \textbf{Automatic layout.} As previously noted the automatic layout feature was sub-optimal and requires further research to improve the display arrangement of the different visual \textit{nodes}. The approach should ideally optimize placement based on the type of artifact rather than considering all the elements on the same level (\textit{e.g.}, it may make sense to tend to represent volumes and networks on the lower half of the layout).
    \item \textbf{Visual feedback.} This is a broad subject, in which we may consider minor changes, such as more detailed status indicators, to more substantial improvements, such as optimizing feedback for other technologies, like Docker Swarm and its multiple containers per service.
    \item \textbf{Static validation.} While the prototype considers validations for some property fields, there is the potential to further enrich this feature with even more. These include validations for ports taking into account the available host ports.
    \item \textbf{Exploring liveness.} \textit{Liveness} is a characteristic of development environments that refers to its ability to provide information to the developer about what they are constructing~\cite{Aguiar2019}. Tanimoto established a scale that can be used to evaluate the level of liveness of an environment~\cite{Tanimoto2013}. 
    
    We may use liveness as an indicator of how much these tools can provide timely feedback to their users and, therefore, how much they can discourage switching between these and other applications. While the previous two points above already contribute towards a more live experience, we can see the possibility of exploring this concept more exhaustively to improve the process of defining orchestrations. 
\end{itemize}

\subsubsection{Empirical Evaluation}

Replication of our study would help to consolidate and increase the confidence level of these results. With this in mind, we have compiled a replication package as described in Section~\ref{sec:procedure}.

Furthermore, we can see different variants to our controlled study that would complement the results presented in this article:

\begin{itemize}
    \item Running the same study with professionals would show how much the results we obtained here with beginners are really extendable to professional software developers. 
    
    \item Conditioning the CG to use a terminal built-into the text editor, instead of an external one, would provide a baseline that is closer to the environment currently used by developers and a more accurate account of the amount of context-switching in question (\textit{cf}. Section~\ref{sec:context-switches}).
    
    \item Expanding our study design to consider \textit{DockStation}, \textit{Admiral} and \textit{Docker Studio} would show the extent to which our findings hold independently of the idiosyncrasies of specific implementations.
    Even though these two tools don't provide a complete visual notation for Docker Compose files, they could still allow narrower studies to be done on the merits of visual programming languages in this domain.
\end{itemize}

While controlled user studies are powerful in identifying isolated cause-effect relations, they fail to fully capture the intricacies of real-world scenarios. Case Conducting a case study with Docker Composer in an industrial environment would be another invaluable source of insight into the actual behavior of the approach in more realistic scenarios.

\section*{Acknowledgments}
Thank you to David Reis, Jessica Diaz, and the participants and anonymous reviewers of the LowCode 2020 workshop~\cite{models2020companion} for discussing earlier versions of this work with us. Also, we thank the anonymous reviewers of the SoSym journal who, through their comments and suggestions, have helped considerably to improve the article's clarity. This work was partially funded by the Integrated Masters in Informatics and Computing Engineering of the Faculty of Engineering, University of Porto (FEUP).

\bibliographystyle{ieeetr}
\bibliography{references}

\end{document}